\documentclass[11pt]{article}
\usepackage[utf8]{inputenc}
\usepackage{lmodern}
\usepackage{amsmath, amssymb, amsthm}
\usepackage{mathtools}

\usepackage{xcolor}
\definecolor{ForestGreen}{rgb}{0.1333,0.5451,0.1333}
\usepackage[colorlinks=true,linkcolor=ForestGreen,citecolor=blue]{hyperref}

\usepackage[margin=1in]{geometry}
\usepackage{graphics}
\usepackage{pifont}
\usepackage{nicefrac}
\usepackage{tikz}
\usepackage{bbm}
\usepackage[T1]{fontenc}

\usetikzlibrary{arrows.meta}
\usepackage{environ}
\usepackage{framed}
\usepackage{url}
\usepackage{algorithm}
\usepackage[algo2e,linesnumbered,lined,ruled,boxed]{algorithm2e}
\usepackage{algpseudocode}
\usepackage[noabbrev,capitalize,nameinlink]{cleveref}
\crefname{equation}{}{}
\usepackage[labelfont=bf]{caption}
\usepackage{cite}
\usepackage{framed}
\usepackage[framemethod=tikz]{mdframed}
\usepackage{appendix}
\usepackage{graphicx}
\usepackage[textsize=tiny]{todonotes}
\usepackage{tcolorbox}
\allowdisplaybreaks[1]

\newcommand\remove[1]{}

\newtheorem{lemma}{Lemma}[section]
\newtheorem{theorem}[lemma]{Theorem}
\newtheorem*{lemma*}{Lemma}
\newtheorem{corollary}[lemma]{Corollary}
\newtheorem*{corollary*}{Corollary}

\theoremstyle{definition}

\newtheorem*{theorem*}{Theorem}
\newtheorem{definition}[lemma]{Definition}

\newtheorem{prob}[lemma]{Problem}
\newtheorem*{rem*}{Remark}

\newcommand\R{\mathbb{R}}

\newcommand\Z{\mathbb{Z}}

\newcommand\E{\mathbb{E}}

\renewcommand{\bf}{\textbf}
\newcommand{\eps}{\varepsilon}
\renewcommand{\O}{\widetilde{O}}
\newcommand{\Ohat}{\widehat{O}}

\renewcommand{\l}{\langle}
\renewcommand{\r}{\rangle}

\newcommand{\assign}{\leftarrow}

\renewcommand{\forall}{\mathrm{\text{ for all }}}

\newcommand{\tg}{\widetilde{\mathbf{g}}}
\newcommand{\tl}{\widetilde{\boldsymbol{\ell}}}

\newcommand{\grad}{\nabla}

\renewcommand{\hat}{\widehat}

\foreach \x in {A,...,Z}{%
	\expandafter\xdef\csname m\x\endcsname{\noexpand\mathbf{\x}}
}

\newif\ifrandom
\randomtrue
\newcommand{\defeq}{\stackrel{\mathrm{\scriptscriptstyle def}}{=}}

\newcommand{\poly}{{\mathrm{poly}}}

\newcommand{\todolater}[1]{}

\newcommand{\wt}{\widetilde}
\newcommand{\diag}{\mathrm{\textbf{diag}}}
\renewcommand{\bar}{\overline}

\newcommand{\yang}[1]{\textbf{\color{red}[Yang: #1]}}
\newcommand{\arun}[1]{\textbf{\color{blue}[Arun: #1]}}
\newcommand{\sidford}[1]{\textbf{\color{green}[Aaron: #1]}}

\newcommand{\ba}{\boldsymbol{a}}
\newcommand{\bc}{\boldsymbol{c}}

\renewcommand{\bf}{\boldsymbol{f}}
\newcommand{\bg}{\boldsymbol{g}}
\newcommand{\bp}{\boldsymbol{p}}

\newcommand{\bu}{\boldsymbol{u}}
\newcommand{\bv}{\boldsymbol{v}}

\newcommand{\bx}{\boldsymbol{x}}

\newcommand{\bell}{\boldsymbol{\ell}}

\newcommand{\bDelta}{\boldsymbol{\Delta}}
\newcommand{\blambda}{\boldsymbol{\lambda}}
\renewcommand{\root}{\mathsf{root}}

\newcommand{\str}{\mathsf{str}}
\newcommand{\wstr}{\widetilde{\str}}
\newcommand{\cW}{\mathcal{W}}
\newcommand{\new}{\mathrm{new}}
\newcommand{\vol}{\mathrm{vol}}
\newcommand{\cS}{\mathcal{S}}
\newcommand{\cC}{\mathcal{C}}
\newcommand{\cG}{\mathcal{G}}
\newcommand{\cT}{\mathcal{T}}
\newcommand{\cF}{\mathcal{F}}

\newcommand{\OPT}{\mathsf{OPT}}

\newcommand{\threshold}{F}
\newcommand{\almostTime}{\widehat{O}}
\renewcommand{\epsilon}{\eps}

\crefname{algocf}{Algorithm}{Algorithms}

\author{
Jan van den Brand \\
Georgia Institute of Technology\\
\texttt{vdbrand@gatech.edu}
	 \and 
Yang P. Liu \\
Stanford University \\
\texttt{yangpliu@stanford.edu}
\and 
Aaron Sidford \\
Stanford University \\
\texttt{sidford@stanford.edu}
}

\begin{document}

\title{Dynamic Maxflow via Dynamic Interior Point Methods}

\clearpage\maketitle
\thispagestyle{empty}

\begin{abstract}
In this paper we provide an algorithm for maintaining a $(1-\epsilon)$-approximate maximum flow in a dynamic, capacitated graph undergoing edge additions. Over a sequence of $m$-additions to an $n$-node graph where every edge has capacity $O(\mathrm{poly}(m))$ our algorithm runs in time $\widehat{O}(m \sqrt{n} \cdot \epsilon^{-1})$.\footnote{Throughout this paper, $\O(\cdot)$ hides $\poly\log(m)$ factors, and $\Ohat(\cdot)$ hides subpolynomial $m^{o(1)}$ factors.} To obtain this result we design dynamic data structures for the more general problem of detecting when the value of the minimum cost circulation in a dynamic graph undergoing edge additions obtains value at most $F$ (exactly) for a given threshold $F$. Over a sequence $m$-additions to an $n$-node graph where every edge has capacity $O(\mathrm{poly}(m))$ and cost $O(\mathrm{poly}(m))$ we solve this thresholded minimum cost flow problem in $\widehat{O}(m \sqrt{n})$. Both of our algorithms succeed with high probability against an adaptive adversary. We obtain these results by dynamizing the recent interior point method used to obtain an almost linear time algorithm for minimum cost flow \cite{CKLPPS22}, and introducing a new dynamic data structure for maintaining minimum ratio cycles in an undirected graph that succeeds with high probability against adaptive adversaries.
\end{abstract}

\newpage
\thispagestyle{empty}
\setcounter{tocdepth}{2}
\tableofcontents

\normalsize
\pagebreak
\pagenumbering{arabic}

\section{Introduction}
\label{sec:intro}

The design and analysis of \emph{dynamic graph algorithms}, i.e.\ dynamic data structures for solving graph problems, is a rich area with a long history. Over the decades there have been a wide range of advances to foundational dynamic graph problems including minimum spanning trees, shortest paths, and bipartite matching (see \cref{sec:intro:related} for the references).

Despite this extensive progress, obtaining dynamic algorithms for flow and cut problems, e.g.\ maximum flow (maxflow), mincost circulation (mincost flow), and $s$-$t$ minumum cut ($st$-mincut), have been more challenging. To the authors' knowledge, the set of results on dynamic maxflow is limited to results for unweighted and incremental graphs \cite{Goranci,GuptaK21} or that have large, $m^{o(1)}$, approximation factors in undirected graphs \cite{CGHPS20,GRST21}. In this work, we study dynamic maxflow in incremental \emph{capacitated, directed} graphs with small $(1-\eps)$, approximation factors.

Two recent lines of research create hope towards overcoming this historic difficulty of obtaining dynamic algorithms for broader classes of flow problems. First, over the past decade there have been multiple running time improvements for solving maxflow (see \cref{sec:intro:related}), culminating in an almost-linear time algorithm for maxflow \cite{CKLPPS22}. Prior to this line of work, the state-of-the-art was $\O(\min\{m^{3/2},mn^{2/3}\})$ \cite{K73,ET75,GR98} and consequently these advances create the hope of much more efficient dynamic flow algorithms, e.g. faster than $\O(\min\{m^{3/2},mn^{2/3}\})$.

A second recent line of work, perhaps with roots in previous work on multiplicative weight updates for flow problems \cite{GK07,Gupta14}, has focused on improved dynamic algorithms for incremental and decremental variants of bipartite-matching and linear programming. These works are particularly intriguing because they show that certain optimization methods such as multiplicative weights or entropy-regularized optimal transport can be used to give dynamic algorithms for partially dynamic bipartite matching and positive linear programming \cite{JJST22,BKS22}. The recent advances in maximum flow all use a different optimization method, namely interior point methods (IPMs), raising the exciting question of whether these recent advances can be dynamized and leveraged to obtain dynamic algorithms for maximum flow an minimum cost flow.

Our main result is an algorithm that uses a dynamic IPM to maintain a $(1-\eps)$-approximate maxflow on incremental capacitated, directed graphs in $\Ohat(m\sqrt{n} \cdot \eps^{-1})$ total time (\cref{thm:inc_maxflow}). Instead of directly solving incremental maxflow, one of the main conceptual contributions of this work is to introduce (and develop efficient algorithms for) a more challenging decision variant of incremental mincost flow which we call \emph{incremental thresholded mincost flow} (see \cref{prob:threshold}). This problem asks to determine when the cost of a mincost flow in an incremental graph is at most a fixed threshold $F$, \emph{exactly}. We show that even on $n$-node $m$-edge graphs with polynomially bounded capacities and costs this problem can be solved exactly, whp.\ in $\almostTime(m \sqrt{n})$ (\cref{thm:inc}).

Our result on dynamic incremental thresholded mincost flow may seem quite surprising in light of the substantial progress over the past few years on showing conditional hardness for closely related dynamic graph algorithms. In particular, it is known that under the online matrix-vector (OMv) conjecture \cite{HenzingerKNS15} that maintaining exact incremental or decremental matching (and hence maxflow) requires total time $\Omega(n^3)$ in dense graphs. Thus our $\almostTime(m \sqrt{n})$ runtime on incremental thresholded mincost flow (\cref{thm:inc}) demonstrates that the decision version is strictly easier than that of maintaining a solution exactly.

Beyond shedding light on the source of hardness in incremental flow and matching problems, our result for incremental threshold mincost flow yields an $\almostTime(m \sqrt{n} \cdot \epsilon^{-1})$ time algorithm for maintaining an $(1-\epsilon)$-approximate maxflow in an $n$-node dynamic graph with polynomial edge capacities undergoing a sequence of $m$-edge additions. This implication follows from a direct reduction of simply applying this dynamic thresholded mincost flow algorithm for thresholds $(1+\eps/2)^i$ for $|i| \le \O(\eps^{-1})$. Interestingly, it is unclear how to use our techniques to obtain this result for approximate incremental maxflow without essentially considering a dynamic thresholded maxflow problem. Also, if we were to solve the incremental thresholded mincost flow problem in $\Ohat(m)$ time (which we are optimistic about, see \cref{sec:barrier}), this would imply a $\Ohat(m \eps^{-1})$ time dynamic incremental maxflow algorithm. Such an $\eps$ dependence would be optimal under OMv, because a better $\eps$ dependence and setting $\eps = 1/n$ would imply an exact dynamic incremental matching algorithm in time $\Ohat(mn^{1-\delta})$.

All our aforementioned results are obtained even against an \emph{adaptive adversary}. Throughout this paper, we use the term \emph{adaptive adversary} to refer to the dynamic graph model where updates can be chosen based on the entire state of the algorithm, including its internal randomness. This is stronger, i.e.\ more difficult to develop algorithms for, than a closely related adaptive adversary model where the dynamic graph updates can be chosen based only on the output of the algorithm. This weaker model is obtained trivially by our algorithms by the fact that for the thresholded problems the output of any algorithm solving the problem is fixed.

To obtain our results we introduce two key tools of intrinsic interest. First, we show that indeed the optimization methods used by \cite{CKLPPS22} in the recent work on almost-linear time maxflow can be dynamized to obtain this result provided a particular data structure for dynamically maintaining a \emph{minimum-ratio cycle} (see \eqref{eq:mrc} in \cref{def:mrc}). Second, we provide data structures for solving minimum-ratio cycle against an adaptive adversary (\cref{thm:minratio})

\paragraph{Paper Overview.} In the remainder of this introduction we formally present our main results in \Cref{sec:intro:results} and discuss related work in \Cref{sec:intro:related}. In \Cref{sec:overview} we then provide a more comprehensive overview of the difficulties in obtaining these results and our approach for overcoming them. Preliminaries are then covered in \Cref{sec:prelim}, our dynamic IPM is covered in \Cref{sec:incrementalipm}, and our dynamic minimum ratio cycle data structure is covered in \Cref{sec:incrementalds}.

\subsection{Results}
\label{sec:intro:results}

The central problem we consider in this paper is the following problem of incremental thresholded mincost flow. This problem is the natural decision variant of dynamic incremental mincost flow. 

\begin{prob}[Incremental thresholded mincost flow]\label{prob:threshold}
	In the \emph{incremental thresholded mincost flow} problem an algorithm is given an empty graph on $n$-vertices $V$ and a \emph{threshold} $\threshold \in \Z$. There are then a sequence of $m$ edge \emph{insertions} where a specified directed edge $e$ is added to the graph's edge set $E$ with an integral cost $\bc_e$ and capacity $\bu_e \ge 0$ each bounded by $\poly(m)$. The algorithm must output after each insertion whether there is a circulation $\bf \in \R^E$ that is feasible, i.e.\  $0 \le \bf_e \le \bu_e$, for all 
	$e \in E$
	and has cost at most $\threshold$, i.e.\ $\bc^\top\bf = \sum_{e \in E} \bc_e\bf_e \le \threshold$.
\end{prob}

Our main result of this paper is an $\almostTime(m \sqrt{n})$ time algorithm with high probability (whp.) against an adaptive adversary as stated in the following theorem.

\begin{theorem}[Incremental mincost flow]
	\label{thm:inc}
	There is an algorithm that solves incremental thresholded mincost flow (\Cref{prob:threshold}) against an adaptive adversary whp.\ in total time $\almostTime(m\sqrt{n})$.
\end{theorem}

\cref{thm:inc} implies corresponding results for solving thresholded $st$-maxflow in incremental graphs using the standard trick of adding an edge from $t \to s$ with sufficiently large capacity and cost $-1$. By setting $\threshold = (1+\eps/2)^i$ for $\O(\eps^{-1})$ values of $i$, we obtain an algorithm for maintaining a $(1-\epsilon)$-approximate maxflow in time $\Ohat(m\sqrt{n} \cdot \eps^{-1})$ for incremental dynamic graphs. Below we define this approximate incremental maxflow problem and provide our results for solving it.

\begin{prob}[Approximate incremental maxflow] \label{prob:maxflow} 
In the \emph{approximate incremental maxflow} problem an algorithm is given a set of $n$ vertices $V$, distinct vertices $s,t \in V$, and an accuracy $\epsilon \in (0,1)$. There are then a sequence of $m$ edge \emph{insertions} where a specified directed edge $e$ is added to the graph's edge set $E$ with an integral capacity $\bu_e \ge 0$ bounded by $\poly(m)$. The algorithm must maintain (after each insertion) an $s$-$t$ flow $\bf \in \R^E$ that is feasible, i.e.\ $0 \le \bf_e \le \bu_e$ for all $e \in E$, and $(1-\epsilon)$-optimal, i.e.\ has value at least $(1 - \epsilon) F^*$, where $F^*$ is the optimal flow value in $(V, E, \bu)$ at that time.
\end{prob}

\begin{theorem}[Incremental approximate maxflow]
	\label{thm:inc_maxflow}
	There is an algorithm that solves approximate incremental maxflow (\Cref{prob:maxflow})  against an adaptive adversary whp.\ in total time $\almostTime(m\sqrt{n} \cdot \epsilon^{-1})$.
\end{theorem}

To the best of our knowledge, no nontrivial algorithm has been developed for this problem previously for weighted graphs, other than \cite{KumarG03} which solves exact incremental weighted maxflow with $n_\Delta^2m$ update time where $n_\Delta$ is the number of vertices affected by the augmenting flow. Na\"{i}vely, the problem can be solved by simply recomputing a maxflow after each addition which using \cite{CKLPPS22} would lead to a runtime of $\Ohat(m^2)$.

Though we focused on the maxflow problem in \Cref{prob:maxflow} and \Cref{thm:inc_maxflow}, it is worth noting that \Cref{thm:inc}, and more broadly, any algorithm for solving incremental thresholded mincost (\Cref{prob:threshold}), can be use to solve a broader range of problems, e.g.\ incremental $st$-shortest path with negative lengths and approximate incremental $st$-mincut. Interesting, the incremental mincost flow problem also encompasses the problem of \emph{incremental cycle detection}. Thus, \cref{thm:inc} matches the runtime of \cite{BC18} up to $m^{o(1)}$ factors, though more recently the total runtime was improved to $\O(m^{4/3})$ \cite{BK20}.

In \Cref{sec:overview} we provide an overview of our approach for obtaining these results. Beyond proving \Cref{thm:inc} and \Cref{thm:inc_maxflow} we provide an additional data struture of possible indpendent interest. In \cref{thm:minratio} we give a data structure that given vectors $\bg \in \R^E$ and $\bell \in \R_{\ge0}^E$ that are dynamically changing, maintains an $m^{o(1)}$-approximate solution to the min-ratio cycle problem $\min_{\mB^\top \bc = 0} \bg^\top\bc / \|\mL\bc\|_1$ against adaptive adversaries in amortized $\Ohat(\sqrt{n})$ time. In comparison, \cite{CKLPPS22} provided an algorithm that ran in $m^{o(1)}$ amortized time, but only worked against oblivious adversaries (which they showed sufficed for the setting of static mincost flow).
\begin{definition}[Min-ratio cycle]
\label{def:mrc}
On a graph $G = (V, E)$, let $\bg \in \R^E$ and $\bell \in \R^E_{\ge0}$. The \emph{minimum-ratio cycle} problem is
\begin{align}
\min_{\mB^\top \bc = 0} \frac{\bg^\top \bc}{\|\mL \bc\|_1} \label{eq:mrc}.
\end{align}
\end{definition}

\subsection{Related work}
\label{sec:intro:related}

\paragraph{Mincost flow via IPMs.} Over the last decade, there have been multiple advances in solving maxflow using a method for linear programming known as \emph{interior point methods} (IPMs). \cite{M13,LS14,M16,CMSV17,LS20,KLS20,AMV20,BLNPSSSW20,BLLSSSW21,GLP21,AMV21,DGGLPSY22,BGJLLPS22,CKLPPS22}. Most of these works used a path-following IPM that ran in $\O(\sqrt{m})$ or $\O(\sqrt{n})$ iterations, where each iteration computed an electrical flow. Our work dynamizes the most recent work of \cite{CKLPPS22} which used a potential-reduction IPM that solves a sequence of dynamic min-ratio cycle problems. We believe that it is an interesting question whether it is possible to efficiently dynamize a path-following IPM.

\paragraph{Dynamic graph algorithms.} The authors only know of a few works on dynamic maxflow.
There is a folklore $O(m)$ time for fully dynamic maxflow on unit-capacitated graph by finding one augmenting flow per step. \cite{KumarG03} maintains dynamic maxflow in incremental graphs $n_\Delta^2m$ time where $n_\Delta$ is number of affected vertices of the augmenting flow. 
\cite{GuptaK21} solves exact incremental unit-capacitated maxflow in $O(n)$ amortized update time, and incremental maximum matching in $O(n)$ amortized update time. The work \cite{Goranci} solves $(1+\eps)$-approximate incremental maxflow in unit-capacity graphs in amortized $\Ohat(\sqrt{m}\eps^{-1/2})$ time, and \cite{CGHPS20,GRST21} maintain a $m^{o(1)}$-approximate maxflow in undirected capacitated graphs.

There is a vast amount of literature on dynamic algorithms for other graph problems such as minimum spanning trees, shortest paths, and matchings.
For maintaining spanning trees and MSTs, there are several results even in the amortized, deterministic, and worst-case settings \cite{EGIN97,Fred85,HK97,HK99,HLT01,HT97,Thorup00,HHKP17,WN13a,HRW15,PD06,WN17,NS17,NSW17,CGLNPS20}. There are several variants of dynamic shortest path problems: we refer the readers to \cite{BGS21} and references therein.

There are previous works maintaining fully dynamic matchings in $n^{1.447}$ time \cite{Sankowski07}, and $n^{1.407}$ time \cite{BrandNS19}.
There are several papers maintaining approximate matchings with an approximation factor of $2$ or larger, eg. via maximal matchings \cite{BaswanaGS15,Solomon16,BhattacharyaHN16,ArarCCSW18,CharikarS18,BernsteinFH21,BhattacharyaK19,BehnezhadDHSS19,ChechikZ19,Wajc20,BhattacharyaK21,Kiss21,LeMSW22}.
For approximation factors between $(1+\eps)$ and $2$, beyond \cite{GuptaP13}, most works study either incremental or decremental settings \cite{PelegS16,BernsteinS15,BernsteinS16,GrandoniSSU22,Kiss21,BehnezhadLM20,Wajc20,BehnezhadK22,RoghaniSW22,AssadiBKL22,BhattacharyaK21,LeMSW22}. Recently, it has been shown that one can maintain better than $2$-approximate matchings in fully dynamic graphs in $\O(1)$ time \cite{BhattacharyaKSW22,Behnezhad22}.
For incremental/decremental graphs with node insertions/deletion on only one side of the bipartite graph, an exact maximum bipartite matching can be maintained in $\tilde{O}(m\sqrt{n})$ total time \cite{BosekLSZ14,BernsteinHR19}.

\paragraph{Conditional hardness of dynamic algorithms.} There are several conditional lower bounds for dynamic algorithms for maxflow and matchings. A particularly popular conjecture is the online matrix-vector (OMv) conjecture. \cite{HenzingerKNS15} introduced the OMv-conjecture and used it to show a $\Omega(\sqrt{m})$ amortized runtime lower bound for incremental and decremental exact bipartite matching. This lower bound was improved to $\Omega(n)$ by \cite{Dahlgaard16}, which also showed a $\Omega(m)$ lower bound on sparse directed weighted graphs for incremental and decremental exact maxflow using OMv.
The work \cite{BrandNS19} showed an amortized $n^{1.407}$ lower bound for exact fully dynamic bipartite matching conditional on the ``hinted uMv-conjecture'' and current bounds on fast matrix multiplication.

Assuming the 3SUM conjecture, \cite{KopelowitzPP16} showed a amortized $n^{1/3}$ time lower bound for incremental exact biparite matching under edge insertions, and $n^{0.39}$ for node insertions. For the problem of fully dynamic exact bipartite matching, \cite{AbboudW14} showed a lower bound of amortized $n^2$ time for combinatorial algorithms and $n^{\omega-1}$ in general based on triangle detection. \cite{AbboudD16} showed a conditional lower bound of amortized $n^{1/2}$ time in planar graphs for fully dynamic maximum weight bipartite matching assuming that no subcubic algorithms exist for general APSP.
\section{Overview}
\label{sec:overview}

Here we describe our approach for solving incremental thresholded mincost flow (\Cref{prob:threshold}), which in turn gives an algorithm for incremental approximate maxflow (\Cref{prob:maxflow}). In \Cref{sec:overview:challenges} we briefly discuss some challenges with alternative approaches to solving these problems, in \cref{overview:incrementalipm} we then describe our optimization outer loop based on a potential reduction IPM, and in \cref{overview:incrementalds} we describe our dynamic data structure that solves approximate min-ratio cycle against adaptive adversaries in amortized $\Ohat(\sqrt{n})$ time per update.

\subsection{Challenges and alternative approaches}
\label{sec:overview:challenges}

Before presenting our approach, let us first consider natural approaches to closely related variants of incremental thresholded mincost flow (\Cref{prob:threshold}) and incremental approximate maxflow (\Cref{prob:maxflow}).
There are two previous works that give algorithms to maintain a $m^{o(1)}$ approximate maxflow in \emph{undirected graphs} by maintaining an $\ell_\infty$ congestion approximator (an object which approximates flows up to $m^{o(1)}$ accuracy) \cite{CGHPS20,GRST21}. However, these methods naturally incur at least polylogarithmic approximation factors due to the quality of congestion approximators \cite{Racke08}, while we want $(1+\eps)$-approximations. Another major issue is that the directed graphs that we study do not admit tree-based approximators to the maxflow. Even in undirected graphs, while it is also known how to boost $\ell_\infty$ congestion approximators to compute $(1+\eps)$-approximate maxflows statically \cite{S13,KLOS14,P16}, it is very unclear that one can implement these procedures dynamically in any way, because every iteration of the algorithms in \cite{S13,KLOS14,P16} require $m^{1+o(1)}$ time. 

The authors know only of a few previous works that study incremental maxflow in \emph{unit-capacitated graphs}. The first is \cite{Goranci}, which achieves a $\Ohat(\sqrt{m}\eps^{-1/2})$ amortized update time for $(1-\eps)$-approximations, or \cite{GuptaK21} which has amortized update time $O(n)$ for exact maxflow. The idea of \cite{Goranci} is to consider two cases. If the current flow is at least $\sqrt{m}\eps^{-1/2}$, then because the graph is unweighted, the algorithm can wait for at least $\eps^{1/2}\sqrt{m}$ edge insertions before performing an almost-linear time recomputation. Otherwise, the algorithm uses an incremental reachability data structure \cite{Italiano86} to augment a path in $O(m)$ time. This approach does not extend readily to capacitated graphs because maintaining an incremental reachability data structure could only us to route a single unit of flow (in the worst case), which maybe insufficient progress on a graph with potentially large capacities.

\subsection{Incremental mincost flow IPM}
\label{overview:incrementalipm}

Given the discussion in \cref{sec:overview:challenges}, it is unclear how to use any combinatorial or optimization-based algorithm to maintain $(1+\eps)$-approximate maxflows in capacitated graphs dynamically. Instead of directly maintaining an approximate maxflow, as discussed in the introduction (\cref{sec:intro}) we pass through the thresholded mincost flow problem (\Cref{prob:threshold}). This interacts very nicely with a potential reduction IPM in the recent work \cite{CKLPPS22}, which we now describe.

As described in \cref{prob:threshold}, our goal is to give an algorithm that for a fixed integer $F \ge 0$, and dynamic graph $G$ undergoing edge insertions of directed edges with costs $\bc_e$ and capacities $\bu_e \ge 0$, decides the first time that a mincost circulation has cost at most $F$. Let us first recall the approach of \cite{CKLPPS22} for solving mincost flow \emph{statically} by using a potential reduction IPM. Define the potential
\begin{align}
\Phi(\bf) \defeq 20m \log(\bc^\top \bf - F) + \sum_{e \in E} (\bf_e+\delta)^{-\alpha} + (\bu_e - \bf_e)^{-\alpha}. \label{eq:potentialoverview}
\end{align}
where $\delta = \frac{1}{20 m^2 C}$ and $\alpha = \frac{1}{5000 \log mCU}$. The choice of $\delta$ is to allow the zero flow $\bf = \vec{0}$ to be strictly feasible. Equivalently, we relax the capacity constraint to $-\delta \le \bf_e \le \bu_e$, and note that this relaxation only changes the optimal cost of a circulation by a small constant (\cref{lemma:delta}).

Thus, $\Phi(\vec{0}) < \infty$. Also, note if there is \emph{no flow} with $\mB^\top \bf = 0$ and $\bc^\top \bf \le F$ then the potential is lower bounded. On the other hand, if such a flow exists, then the potential is unbounded: results of \cite{CKLPPS22} show that by solving a \emph{min-ratio cycle problem} we can compute a step $\bDelta$ such that the potential of the flow decreases by at least $-\wt{\Omega}(1)$ (\cref{lemma:quality,lemma:phidecrease}), i.e. $\Phi(\bf + \bDelta) - \Phi(\bf) \le -\wt{\Omega}(1).$ By repeated application, the potential can be made arbitrarily small.

Precisely, the min-ratio cycle problem asks to report an $m^{o(1)}$ approximation to the minimizer of the problem in \eqref{eq:mrc}:
\begin{align*} \min_{\mB^\top \bDelta = 0} \frac{\bg^\top \bDelta}{\|\mL \bDelta\|_1}, \end{align*} where $\bg \defeq \nabla \Phi(\bf)$ is the \emph{gradient} of $\Phi$ and $\mL = \diag(\bell)$ for $\bell_e \defeq (\bf_e + \delta)^{-1} + (\bu_e - \bf_e)^{-1}$ are the \emph{lengths}.
If the objective in \eqref{eq:mrc} has value at most $-m^{-o(1)}$, then updating $\bf \assign \bf + \eta\bDelta$ for a proper choice of scaling $\eta$ decreases $\Phi(\bf)$ by at least $m^{-o(1)}$ (\cref{lemma:phidecrease}). Otherwise, the algorithm can actually conclude that there is no flow with $\mB^\top \bf = 0$ and $\bc^\top \bf \le F$, so our data structure can advance to the next edge insertion.
By running $O(m^{1+o(1)} \poly(\log mCU))$ steps, the cost of $\bf$ satisfies $\bc^\top \bf \le F + (mCU)^{-10}$ (\cref{lemma:ending}), so our algorithm can terminate successfully. Finally, the algorithm still succeeds if $\bg$ and $\mL$ are replaced with approximations $\bg \approx \wt{\bg}$ and $\mL \approx \wt{\mL}$. Over the course of the whole algorithm, entries $\bg_e$ and $\mL_e$ over all edges $e$ change at most $m^{1+o(1)}$ times.

To finish our overview of the IPM analysis, we discuss how the potential changes under edge insertions. First, note that if an edge $e$ is inserted, then we can set $\bf_e = 0$ without changing the demand of $\bf$, and without breaking feasibility (because $-\delta \le 0 \le \bu_e$). Let us understand how $\Phi(\bf)$ changes. Note that almost all terms in \eqref{eq:potentialoverview} stay the same, except there is a new term corresponding to edge $e$ which has value $(\bf_e + \delta)^{-\alpha} + (\bu_e - \bf_e)^{-\alpha} = \delta^{-\alpha} + \bu_e^{-\alpha} \le O(1)$ by the choice of $\alpha$ (\cref{lemma:ipminsert}). Hence, over the course of $m$ edge insertions, the potential increases by a total of $O(m)$. In each step of the IPM, assuming there is currently a feasible flow with cost at most $F$, the potential decreases by $m^{-o(1)}$. So the total number of iterations is still $m^{1+o(1)}$ as desired.

To summarize, if there exists a feasible flow $\bf$ with $\bc^\top\bf \le F$, then solving a min-ratio cycle problem decreases the potential by $-\wt{\Omega}(1)$. Thus, if the value of the min-ratio cycle problem is not negative enough, then we can conclude that no feasible flow $\bf$ has $\bc^\top\bf \le F$. Each edge insertion increases the potential by $O(1)$, which sums to $O(m)$ over the whole algorithm.
We note that it is unclear how to perform such an analysis with an IPM (either path-following or primal-dual potential reduction) that has $\O(\sqrt{m})$ iterations, and believe that this is an interesting conceptual direction to study.

\subsection{Dynamic minimum ratio cycle}
\label{overview:incrementalds}

The potential reduction IPM solves mincost flow via a sequence of min-ratio cycle problems, for which \cite{CKLPPS22} provides a data structure with amortized $m^{o(1)}$ time per update in the case of static maxflow. In the dynamic setting, for technical reasons discussed in \cref{sec:barrier}, we adapt the min-ratio cycle data structure of \cite{CKLPPS22} to succeed against adaptive adversaries in general, at the cost of increasing the amortized runtime to $\O(\sqrt{n})$. A more na\"{i}ve modification would result in amortized runtime $\Ohat(\sqrt{m})$, so we use an additional trick (discussed at the en of this section) of layering a chain of spanners on top of the data structure to reduce $m$ to $n$.

Throughout our description, we refer to the ratio $\bg^\top \bc / \|\mL \bc\|_1$ as the \emph{quality} of the cycle $\bc$. Our goal is to find a cycle of sufficiently good quality. The construction is based on the dynamic min-ratio cycle data structure of \cite{CKLPPS22}, modified to work against adaptive adversaries. At a high level, the data structure involves a vertex reduction prodecure, an edge sparsification procedure, and careful recursive application of these procedures. We let $k$ be a size reduction parameter which we will set as $k = m^{o(1)}$.

\paragraph{Vertex reduction.} The vertex reduction procedure takes the original graph $G$, and outputs $t = \O(k)$ graphs which are \emph{contractions} of $G$, such that on average, the quality of the min-ratio cycle has only gotten worse by a $\O(1)$ factor. Precisely, the algorithm dynamically maintains $\O(k)$ forests $F_1, \dots, F_t \subseteq E(G)$, each of which has $O(m/k)$ connected components, such that on average over $i$, the stretch of edge $e$ (ratio of the length of the path between the endpoints of $e$ in $F_i$ to its length) is $\O(1)$ for all edges $e$ (\cref{lemma:dynamiclsd}). This allows us to prove that for at least half of the contracted graphs $G / F_1, \dots, G / F_t$, the quality of the best min-ratio cycle has not gotten worse by more than a $\O(1)$ factor. The algorithm dynamically maintains all the graphs $G / F_1, \dots, G / F_t$, in order to succeed against an adaptive adversary. This is the key difference from the algorithm of \cite{CKLPPS22}: in the setting of static mincost flow, they prove that using properties of the IPM, they can randomly sample $O(\log m)$ graphs out of $G / F_i$ and only maintain these. As discussed in \cref{sec:barrier}, we cannot show that this works in our setting.

\paragraph{Edge reduction via spanners.} The graphs $G / F_i$ still have $\Omega(m)$ edges, despite have $O(m/k)$ vertices, as each $F_i$ has $O(m/k)$ connected components. Thus, we will dynamically maintain a spanner on each graph $G / F_i$. Over the course of edge insertions to $G$, the forest $F_i$ is decremental (see \cref{lemma:dynamiclsd}). When an edge of $F_i$ is deleted, a connected component in the forest $F_i$ splits into two, so the graph $G / F_i$ undergoes \emph{vertex splits}, where a vertex of $G / F_i$ becomes two vertices.
So we must maintain a spanner of $G / F_i$ under edge insertions, deletions, and vertex splits. Fortunately, such an object was constructed for us in \cite{CKLPPS22}, with amortized runtime $\Ohat(k)$ and recourse $\Ohat(1)$, where recourse means the number of changes to the spanner per change to $G / F_i$.

\paragraph{A recursive data structure.} Our data structure alternately uses vertex reduction and edge reduction data structures for $d$ layers. Thus, at the bottom layer, the graphs will have about $m/k^d$ edges, and there will be $\O(k)^d$ graphs total. This is because each $G / F_i$ has $O(m/k)$ vertices, and hence has spanner with $\Ohat(m/k)$ edges. Also, there are $\O(k)$ forests $F_i$, so each layer increases the number of graphs by a factor of $\O(k)$. Consider setting $d$ so that $k^d \approx \sqrt{m}$: this way, there are $\Ohat(\sqrt{m})$ graphs at the bottom level, each with about $\Ohat(\sqrt{m})$ edges. The formal algorithm performing this corresponds to \cref{lemma:dynamicchain}.

\paragraph{Querying for a cycle.} We now describe how to extract an approximate min-ratio cycle out of the data structure we have described. There are two main types of cycles we look for. Consider a graph $G / F_i$ and a sparsifier $H \subseteq G/F_i$. Consider an edge $e \in E(G/F_i) \setminus E(H)$, and let $\Pi(e)$ be a short path in $E(H)$ between the endpoints of $e$. Then the concatenation $e \oplus \Pi(e)$ is a cycle: we call these \emph{sparsifier cycles} (\cref{def:sparsifiercycle}). As in \cite{CKLPPS22}, we show that either there is some sparsifier cycle that we can return (and whose quality is sufficient), or $H$ supports a cycle with good quality (see \cref{sec:query}).

The dynamic spanner data structure explicitly maintains all the paths $\Pi(e)$ (\cref{thm:spanner}), and hence we can maintain the quality of all cycles $e \oplus \Pi(e)$ in constant overhead. Thus, it remains to analyze the case where no cycle $e \oplus \Pi(e)$ has good enough quality. In that case, we prove that a uniformly random graph among the $\O(k)^d$ graphs at the $d$-th level of the data structure supports a good quality cycle with probability at least $1/2$ (see \eqref{eq:dsstretch}). Thus, we can randomly sample $O(\log m)$ graphs at the bottom level of the data structure (each with $\Ohat(\sqrt{m})$ edges). On each of these graphs, we statically compute an approximately optimal min-ratio cycle in $\Ohat(\sqrt{m})$ time using any of a variety of methods (for example, just running our data structure again, or building an $\ell_1$ oblivious routing). Because we sample $O(\log m)$ graphs, at least one has a good quality cycle whp.

\paragraph{Reducing $m$ to $n$.} In \cref{thm:minratio} we achieve an amortized runtime of $\Ohat(\sqrt{n})$ instead of $\Ohat(\sqrt{m})$ as described in this overview. To achieve the tighter amortized runtime, we apply the dynamic spanner data structure to our original graph until it is down to $\Ohat(n)$ edges. As a technical point, this has to be done in several layers, instead of one shot, because our data structure pays an amortized runtime of $\Ohat(k)$ (even though the recourse is still $\Ohat(1)$) for reducing the number of edges by a factor of $k$, so we cannot afford to set $k = m/n$. This reduction is formally described in the definition of a \emph{spanner chain} (\cref{def:spannerchain}).

\subsection{Difficulties towards optimal runtimes}
\label{sec:barrier}
The authors suspect that (with further insights) it may be possible to improve \cref{thm:inc} and solve thresholded mincost flow with total runtime $\Ohat(m)$ (at least against an oblivious adversary that cannot choose the edge additions as a function of the algorithm's internal randomness).  To understand the main technical hurdle to achieving such a result, we must describe a technical point in the data structure of \cite{CKLPPS22} for min-ratio cycle. In the approach of \cite{CKLPPS22}, a simple version of the min-ratio cycle data structure has the following issue: sometimes the data structure will return that it cannot find a min-ratio cycle of the desired quality, even though we know that one exists (by the guarantees of the outer IPM loop). In this case, \cite{CKLPPS22} proved that there is a strategy for rebuilding some levels of their recursive data structure to ensure that a good cycle is found in a low \emph{amortized} amount of time.

The reason this does not extend to our setting is because there is another possible reason that we cannot find a cycle of sufficiently good quality: it actually does not exist, because there is no flow of cost at most the desired amount $F$. So in some sense, we would not know whether we need to rebuild data structure levels, or simply move on to the next iteration (because we have confirmed that our graph does not have a flow of cost at most $F$).

Finally, we note that while achieving an $\Ohat(m)$ time algorithm for incremental maxflow against oblivious adversaries is a natural next step, it is much more unclear to us whether there is a natural direction to achieve improvements over our work against adaptive adversaries.

\section{Preliminaries}
\label{sec:prelim}

\paragraph{General notation.} We use $G = (V, E)$ to refer to a graph with vertices $V$ and edges $E$ and we let $V(G)$ and $E(G)$ denote the vertex set and edge set of a graph $G$. Throughout, $\deg(v)$ refers to the number of adjacent edges to $v$. Given a vector $\bx$, we let $|\bx|$ denote the coordinate-wise absolute value.
Parameter $m$ specifies the number of edge insertions, which is an upper bound on (but not necessarily identical to) the number of edges in $G$.
For a graph $G$ and an edge subset $F \subseteq E(G)$, we let $G/F$ denote the graph where all edges in $F$ are contracted. For an edge $e \in G$, we let the image of $e$ in $G/F$ be denoted as $\hat{e}$. Note that $\hat{e}$ may be a self-loop. For positive real numbers $a, b$ and $\gamma \ge 1$ we write $a \approx_{\gamma} b$ to mean $\gamma^{-1}a \le b \le \gamma a$.

\paragraph{Dynamic graphs.} A dynamic graph formally is a sequence of graphs, where each graph is updated from the previous via edge insertions and edge deletions. For simplicity, we will denote our dynamic graph simply as $G$. We say that a dynamic graph is \emph{incremental} if there are only edge insertions, and \emph{decremental} if there are only edge deletions.

The graphs in this paper undergo edge insertions, deletions, and \emph{vertex splits}. In a vertex split operation, a vertex $v$ is split into two vertices $v_1$ and $v_2$. The edges adjacent to $v$ are separated amongst $v_1$ and $v_2$, and we assume that such an update is described using $\min\{\deg(v_1), \deg(v_2)\}$ integers, i.e. the edges adjacent to vertex among $v_1$ and $v_2$ of smaller degree. Vertices $v_1$ and $v_2$ are not necessarily connected by an edge, but this may happen if $v$ originally has self-loops.

Throughout, we study data structures that maintain subgraphs $H$ of a dynamic graph $G$. We say that the \emph{recourse} of such a dynamic graph data structure is the amortized number of edge changes to $H$ per edge update to $G$.

\paragraph{Trees and routings.} Given a forest $F$ and two vertices $u, v \in V$ we let $\bp(F[u, v]) \in \R^E$ denote the signed indicator vector representing the path from $u$ to $v$ along $F$. The sign of an edge $+1$ if it is in the direction of the (arbitrary) orientation of the edges of $G$, and $-1$ otherwise. $\bp(F[u, v])$ is undefined if $u, v$ are not in the same connected component of $F$.
\section{Incremental mincost flow IPM}
\label{sec:incrementalipm}

The goal of this section is to provide a potential-reduction IPM that solves incremental thresholded mincost flow (\cref{prob:threshold}). The key conceptual insight is that the IPM outer loop of \cite{CKLPPS22} can readily be dynamized. We will state the main guarantees of the framework, and many proofs and lemmas can be directly cited from \cite{CKLPPS22}.

For the remainder of the section, fix the parameter $\alpha \defeq \frac{1}{5000 \log(mCU)}$.

\begin{algorithm2e}[t!]
	\caption{Incremental mincost flow IPM \label{alg:incremental:ipm}}
	\SetKwProg{Proc}{procedure}{}{}
	\SetKwProg{Params}{parameters}{}{}
	\Params{}{
	$m$: upper bound on number of edge insertions.\\
	$\alpha = \frac{1}{5000 \log(mCU)}$, $\kappa\in(0,1)$.
	}
	\Proc{$\textsc{IncrementalIPM}(G=(V,E), \bc \in \R^E, \bu\in\R^E, F\in\R)$}{
		Let $t=1$, $\bf^{(t)}=0$\\
		\While{$\bc^\top \bf > F+(mCU)^{-10}$}{
			Let $\tg,\tl \in \R^E$ such that $\|\mL_G(\bf^{(t)})^{-1}(\wt{\bg} - \bg_G(\bf^{(t)}))\|_\infty \le \kappa\alpha/32$ and $\wt{\bell} \approx_2 \bell_G(\bf^{(t)})$ 
			\label{line:incremental:find_flow_loop}\\
			Attempt to find $\bDelta\in\R^E$ satisfying $\mB^\top\bDelta = 0$ and $\wt{\bg}^\top\bDelta/\|\wt{\mL}\bDelta\|_1 \le -\kappa\alpha/4$\\
			\If{$\wt{\bg}^\top\bDelta/\|\wt{\mL}\bDelta\|_1 > -\alpha/4$ for all circulations $\bDelta$}{
				Notify user that there is currently no circulation of cost at most $F$.\\
				Then wait for the next edge insertion.\\
				For new edge $e$ let $\bf^{(t)}_e = 0$.\\
				Go back to \Cref{line:incremental:find_flow_loop}	
			}
			$\bf^{(t+1)} \assign \bf^{(t)} + \eta\bDelta$ for $\eta \assign (\kappa\alpha)^2/(16\cdot50|\wt{\bg}^\top\bDelta|)$.\\
			$t \leftarrow t+1$.
		}
		\Return \label{line:ipm:return} $\bf^{(t)}$ which is a flow of cost at most $F+(mCU)^{-10}$.
	}
\end{algorithm2e}

\begin{theorem}[Incremental IPM guarantees]
\label{thm:incipm}
Consider a incremental thresholded mincost circulation problem on a graph with costs bounded by $C$ in absolute value and capacities bounded by $U$ as described in \cref{prob:threshold}.
For any $\kappa \in (0, 1)$, \Cref{alg:incremental:ipm} satisfies the following:
\begin{enumerate}
\item After initialization and after each edge insertion, the algorithm either (i) correctly identifies if there is no circulation of cost at most $F$, or (ii) returns a circulation $\bf^{(t)}$ with cost at most $F+(mCU)^{-10}$.
\item When a circulation $\bf^{(t)}$ is returned on \Cref{line:ipm:return}, we have $t\le\O(m(\alpha\kappa)^{-2}\log CU)$.
\end{enumerate}
\end{theorem}

Ultimately, we will also prove that we can ensure that $\wt{\bg}$ and $\wt{\bell}$ 
have at most  $\Ohat(m\kappa^{-2})$ coordinate changes in total over the whole algorithm.
However, this will require a few details pertaining to the data structure implementation, so we defer the precise method to achieve this to \cref{sec:gl}.

Let the total number of edge insertions be $m$. For a graph $G$ with edges with capacities bounded by $U \le \poly(m)$, costs $\bc$ bounded by $C \le \poly(m)$, and target total cost $F$, define the potential function $\Phi$ on flows $\bf \in \R^{E(G)}$ as
\begin{align}
\Phi_G(\bf) \defeq 20m \log(\bc^\top \bf - F) + \sum_{e \in E(G)} (\bu_e - \bf_e)^{-\alpha} + (\bf_e + \delta)^{-\alpha}, \label{eq:phi}
\end{align}
for $\alpha = \frac{1}{5000 \log(mCU)}$ and $\delta = \frac{1}{20m^2C}$. By decreasing the potential $\Phi_G$, the algorithm trades off between decreasing the total cost $\bc^\top \bf$, and moving $\bf_e$ away from the capacity constraints $0 \le \bf_e \le \bu_e$. As discussed below \eqref{eq:potentialoverview}, the reason for the term $(\bf_e + \delta)^{-\alpha}$ as opposed to the more obvious $\bf_e^{-\alpha}$ is to ensure that the potential is not infinite when we set all $\bf_e = 0$, corresponding to the zero circulation. This technically weakens the capacity constraint on $\bf_e$ to $-\delta \le \bf_e \le \bu_e$, but this does not affect the optimal cost by more than $\delta m^2 C \le 1/20$ (see \cref{lemma:delta}). Further, since $F$ is integral, this looser condition does not affect the first instance where some feasible circulation $\bf^*$ satisfies $\bc^\top \bf^* \le F$.

\begin{lemma}
\label{lemma:delta}
Consider graph $G$ with capacities $\bu_e$ and costs $\bc_e$ with $|\bc_e| \leq C$ for $e \in E$. For $\delta \ge 0$, define \[ \OPT \defeq \min_{\substack{\mB^\top\bf = 0 \\ 0 \le \bf_e \le \bu_e \forall e \in E}} \bc^\top \bf \enspace \text{ and } \enspace \overline{\OPT} \defeq \min_{\substack{\mB^\top\bf = 0 \\ -\delta \le \bf_e \le \bu_e \forall e \in E}} \bc^\top \bf. \]
Then $\overline{\OPT} - \OPT \ge -m^2 C\delta$.
\end{lemma}
\begin{proof}[Proof of \cref{lemma:delta}]
	Let $\bf$ satisfy $\mB^\top\bf = 0$ and $ -\delta \le \bf_e \le \bu_e$ for all $e \in E$.  Further, let $e\in E$ be an edge with $\bf_e < 0$.
	Since $\bf$ is a circulation, it has a cycle decomposition that induces a linear combination $\bf = \sum_i \lambda_i \bc_i$ where each $\bc_i \in \{0,1\}^E$ is a cycle and $\lambda_i \in \R$. 
	By removing all cycles that contain edge $e$ and have negative $\lambda_i$ from this linear combination, the flow on $\bf_e$ becomes non-negative and no previously positive $\bf_{e'}$ edge becomes negative.
	 The cost of the flow might increase by up to $|\bf_e| m C \le \delta m C$.
	As there are at most $m$ edges with negative flow, removing all negative cost edges can change the cost by at most $m^2C\delta$.
\end{proof}

Once the potential decreases to $-\O(m)$, we know that the flow has cost close to $F$.
\begin{lemma}
\label{lemma:ending}
If $\Phi_G(\bf) \le -200 m \log (mCU)$, then $\bc^\top \bf \le F + (mCU)^{-10}$.
\end{lemma}
\begin{proof}
Observe that we always have $\sum_{e \in E(G)} (\bu_e - \bf_e)^{-\alpha} + (\bf_e + \delta)^{-\alpha} \ge 0$ because the flow stays feasible throughout the method (or else the potential would be infinite). Thus $\Phi_G(\bf) \le -200 m \log (mCU)$ implies $20m \log(\bc^\top \bf - F) \le -200m \log(mCU)$ and so $\bc^\top \bf \le F + (mCU)^{-10}$.
\end{proof}

As in \cite{CKLPPS22}, our method is an IPM based on the $\ell_1$ norm to decrease $\Phi_G$. 
To this end, it is useful to define the \emph{gradients} and \emph{lengths} of each IPM subproblem.
\begin{definition}[Gradients and lengths]
\label{def:gl}
On a graph $G$ and flow $\bf \in \R^{E(G)}$ we define the \emph{lengths}
\begin{align}
\bell_G(\bf)_e \defeq (\bu_e - \bf_e)^{-1-\alpha} + (\bf_e + \delta)^{-1-\alpha}
\end{align}
and \emph{gradient}
\begin{align}
\bg_G(\bf)_e \defeq [\grad \Phi_G(\bf)]_e \defeq 20m(\bc^\top \bf - F)^{-1} \bc_e + \alpha(\bu_e - \bf_e)^{-1-\alpha} - \alpha(\bf_e + \delta)^{-1-\alpha}. \label{def:g}
\end{align}
\end{definition}
In each iteration of Algorithm \ref{alg:incremental:ipm}, line \ref{line:incremental:find_flow_loop} defines the $\ell_1$ subproblem \[ \min_{\mB^\top\bDelta = 0} \bg_G(\bf)^\top\bDelta/\|\mL_G(\bf)\bDelta\|_1. \] For sake of efficient implementation, we will study this objective with only approximate gradients and lengths. To analyze such an algorithm, we require the following two statements from \cite{CKLPPS22}. The first says that if there exists a feasible circulation $\bf^* \in \R^{E(G)}$ with $\bc^\top \bf^* \le F$, then the value of the natural $\ell_1$ objective is at most $-{\Theta}(\alpha)$.
\begin{lemma}[\!\!{\cite[Lemma 4.7]{CKLPPS22}}]
\label{lemma:quality}\label{lem:incremental:find_augmentation}
Let $G$ be a graph such that there is a feasible circulation $\bf^*$ with $\bc^\top \bf^* \le F$. Let $\wt{\bg} \in \R^{E(G)}$ satisfy $\left\|\mL_G(\bg)^{-1}\left(\wt{\bg} - \bg_G(\bf)\right) \right\|_\infty \le \eps$ for some $\eps < \alpha/2$, and $\wt{\bell} \in \R^E_{>0}$ satisfy $\wt{\bell} \approx_2 \bell_G(\bf)$. If $\Phi_G(\bf) \le 200m\log mCU$ and $\log(\bc^\top\bf - F) \ge -10 \log mCU$,
then
\[ \frac{\wt{\bg}^\top(\bf^* - \bf)}{\left\|\wt{\mL}(\bf^* - \bf)\right\|_1} \le -\alpha/4. \]
\end{lemma}
Technically, in \cite{CKLPPS22} \cref{lemma:quality} was shown with the $\log mCU$ terms replaced with $\log mU$ where $U$ was an upper bound for both capacities and costs, but the proof is identical.

The second statement says that finding a solution of good quality allows us to significantly decrease the potential.
\begin{lemma}[\!\!{\cite[Lemma 4.4]{CKLPPS22}}]
\label{lemma:phidecrease}\label{lem:incremental:potential_decrease}
Let $\wt{\bg} \in \R^{E(G)}$ satisfy $\left\|\mL_G(\bf)^{-1}\left(\wt{\bg} - \bg_G(\bf)\right) \right\|_\infty \le \kappa'/8$ for some $\kappa' \in (0,1)$, and $\wt{\bell} \in \R^E_{>0}$ satisfy $\wt{\bell} \approx_2 \bell_G(\bf)$.
Let $\bDelta$ satisfy $\mB^\top\bDelta = 0$ and $\nicefrac{\wt{\bg}^\top \bDelta}{\left\|\wt{\mL}\bDelta\right\|_1} \le -\kappa'$.
Let $\eta$ satisfy $\eta \wt{\bg}^\top\bDelta  = -\kappa'^2/50.$ Then $\bf + \eta\bDelta$ is feasible, and
\[ \Phi_G(\bf + \eta\bDelta) \le \Phi_G(\bf) - \frac{\kappa'^2}{500}. \]
\end{lemma}
Our analysis only requires one more simple lemma to handle edge insertions: when an edge is inserted, the potential only increases by a constant.
\begin{lemma}[Edge insertion]
\label{lemma:ipminsert}\label{lem:incremental:edge_insertion}
Let $G$ be a graph, and let $G'$ be a graph with one edge $e^{\new}$ inserted to $G$. Consider a flow $\bf \in \R^{E(G)}$, and extend $\bf$ to $\R^{E(G')}$ by setting $\bf_{e^{\new}} = 0$, so that $\bf$ routes the same demand in $G$ and $G'$. Then $\Phi_{G'}(\bf) - \Phi_G(\bf) \le 3.$
\end{lemma}
\begin{proof}
By definition (see \eqref{eq:phi}) we know that $\Phi_{G'}(\bf) - \Phi_G(\bf) = \bu_{e^{\new}}^{-\alpha} + \delta^{-\alpha} \le 1 + 2 = 3$, because $\bu_{e^{\new}} \ge 1$, and the definition of $\alpha = \frac{1}{5000 \log(mCU)}$ and $\delta = \frac{1}{20m^2C}$.
\end{proof}
We have all the necessary pieces to show \cref{thm:incipm}.
\begin{proof}[Proof of \cref{thm:incipm}]
	We first argue that throughout \Cref{alg:incremental:ipm}, we always have $\Phi_G(\bf^{(t)}) \le 200m\log(mCU)$ as that is a requirement to use \Cref{lem:incremental:find_augmentation}.
	\paragraph{Bounding the potential.}
	At the start of the algorithm, we have $\bf^{(t)} = 0$, so
	$$
	\Phi_G(\bf^{(t)}) = 20m \log(-F) + \sum_{e\in E} (\bu_e^{-\alpha} + \delta^{-\alpha})
	\le
	20m \log(-F) + 3|E|
	<
	100m \log(mCU)
	$$
	where the last step assumes that without loss of generality $|F| \le mCU$. 
	This holds because $m$ is an upper bound on the edges we will insert into the graph $G$ so it would be impossible to find a flow of cost $F$ with $|F| > mCU$.
	
	The potential increases by $(\bu_e^{-\alpha} + \delta^{-\alpha}) \le 3$ whenever we add a new edge (\Cref{lem:incremental:edge_insertion}). 
	Thus $\Phi_G(\bf^{(t)}) \le 200 m \log(mCU)$ because we add at most $3m$ in total to $\Phi_G$.
	At last, note that augmenting $\bf^{(t)}$ by $\bDelta$ will only decrease the potential by \Cref{lem:incremental:potential_decrease}.
	
	In summary, we always have $\Phi_G(\bf^{(t)}) \le 200 m \log(mCU)$.
	
	\paragraph{Correctness of the output.}
	There are two types of output: (i) the IPM claims no circulation of cost at most $F$ exists, then pauses to wait for the next edge insertion, (ii) the IPM terminates, returns $\bf^{(t)}$, and claims this flow has cost at most $F+(mCU)^{-10}$.
	
	The correctness of (ii) is given by the break condition of the While-loop in \Cref{alg:incremental:ipm}.
	
	The correctness of (i) is given by \Cref{lem:incremental:find_augmentation}.
	We argue via \Cref{lem:incremental:find_augmentation} that if there was some circulation of cost at most $F$, then we could find an augmenting flow $\bDelta$.
	The IPM only claims that no circulation of cost at most $F$ exists, if it did not find an augmenting flow, so the output is correct.
	
	To apply \Cref{lem:incremental:find_augmentation} we need to satisfy the following:
	\begin{itemize}
		\item $\wt{\bg} \in \R^{E(G)}$ with $\left\|\mL_G(\bg)^{-1}\left(\wt{\bg} - \bg_G(\bf)\right) \right\|_\infty \le \kappa\alpha/32 < \alpha/2$, and $\wt{\bell} \in \R^E_{>0}$ with $\wt{\bell} \approx_2 \bell_G(\bf)$
		are given by \Cref{line:incremental:find_flow_loop}. 
		\item $\Phi_G(\bf) \le 200m\log mCU$ is given by the previous paragraph.
		\item $\log(\bc^\top\bf - F) \ge -10 \log mU$ is given by (i), i.e.~the condition of our While-loop.
	\end{itemize}
	So by \Cref{lem:incremental:find_augmentation} we have
	\[ \frac{\wt{\bg}^\top(\bf^* - \bf)}{\left\|\wt{\mL}(\bf^* - \bf)\right\|_1} \le -\alpha/4 \]
	which means there exists some augmenting circulation $\bDelta$.
	
	\paragraph{Number of iterations}
	The number of iterations (i.e.~number of augmentations performed to $\bf^{(t)}$) can be bounded by analyzing the potential function $\Phi_G(\bf^{(t)})$.
	By \Cref{lem:incremental:potential_decrease} and $\kappa'=\alpha\kappa/4$, the potential decreases by at least $\kappa'^2/500$ when we augment by $\bDelta$.
	So the potential decreases by $\Omega(\alpha^2\kappa^2)$.
	The value of $\Phi_G(\bf^{(1)}) \le 100m \log(mCU)$ and the potential increases by at most an additional $3m$ over all $m$ edge insertions by \Cref{lem:incremental:edge_insertion}.
	By \Cref{lemma:ending}, if $\Phi_G(\bf^{(1)}) \le -200 m \log (mCU)$ then $\bf^{(t)}$ is a circulation of cost at most $F+(mCU)^{-10}$.
	Thus is takes at most $\tilde{O}(m/(\alpha^2\kappa^2) \log (CU))$ augmentations until the IPM terminates.
\end{proof}
\section{Incremental mincost flow data structure}
\label{sec:incrementalds}

The goal of this section is to give a data structure for dynamic min-ratio cycle against adaptive adversaries, and to use it to dynamically solve mincost flow on incremental graphs.
\begin{theorem}[Dynamic min-ratio cycle]
\label{thm:minratio}
There is a randomized algorithm that on a dynamic graph $G$ with at most $m$ edges at all times, supports the following operations, where $-\O(1) \le \log |\bell_e| \le \O(1)$ for all edges at all times.
\begin{itemize}
\item $\textsc{Insert}(e, \bg_e, \bell_e)$: Insert an edge $e$ with gradient $\bg_e$, and length $\bell_e$.
\item $\textsc{Delete}(e)$: Delete edge $e$.
\item $\textsc{Query}$: Returns a cycle $\bDelta$ satisfying for $\kappa = \exp(-O(\log^{7/8} m \log \log m))$,
\[ \frac{\bg^\top\bDelta}{\|\mL \bDelta\|_1} \le \kappa \min_{\mB^\top\bc=0} \frac{\bg^\top\bc}{\|\mL \bc\|_1}. \]
\end{itemize}
The data structure explicitly maintains a collection of forests $F_1, \dots, F_s \subseteq G$ for $s \le \Ohat(\sqrt{n})$. 
The cycle $\bDelta$ returned by \textsc{Query} is returned 
implicitly by specifying and index $t \in [s]$, and via $L \le \Ohat(1)$ off-tree edges $(u_i,v_i) \in E$ for $i \in [L]$ such that the cycle is \[ ((u_1,v_1), v_1 \to u_2, (u_2, v_2), \dots, (u_L, v_L), v_L \to u_1) \] where ``$v_i\to u_{i+1}$'' denotes the unique path in tree $F_t$ for all $i \in [L - 1]$.
The amortized runtime of each operation is $\Ohat(\sqrt{n})$, and the algorithm succeeds whp. against adaptive adversaries.
\end{theorem}
Throughout the section, we will refer to the vector $\bg$ as the \emph{gradient vector} due to its connection with the IPM, even though $\bg$ can be an arbitrary real vector in the context of \cref{thm:minratio}.

\subsection{Data structure definitions}
\label{sec:dsdef}
In this section we introduce the definitions that make up our main data structure construction. Later in \cref{sec:runtimeofds} we analyze the runtime for maintaining the defined data structures.

Our data structures (as in \cite{CKLPPS22}) are recursive, interlacing partial tree routings and spanners. The spanner reduces the number of edges and maintains the value of the min-ratio cycle problem up to a $\Ohat(1)$ factor, and the dynamic partial tree routing reduces the min-ratio cycle problem to a graph with less vertices (but an equal number of edges), up to a $\O(1)$ approximation factor. We start by defining what it means to maintain a partial tree routing. Our definitions are adapted and in part copied from \cite{CKLPPS22}. To start, we introduce a \emph{rooted spanning forest}.
\begin{definition}[Rooted spanning forest]
\label{def:forest}
A \emph{rooted spanning forest} of a graph $G = (V, E)$ is a forest $F=(V,E')$ with $E'\subset E$
such that each connected component of $F$ has a unique distinguished vertex known as the \emph{root}. 
For a vertex $v \in V$, we denote the root of the connected component that $v$ is in as $\root^F_v$.
\end{definition}
In general, our algorithm will maintain a rooted spanning forest $F$ that undergoes edge deletions. We let $T$ be a tree with vertex set $V$ containing $F$, so $F \subseteq T$, and $T$ does not change under edge deletions to $F$.
After an edge deletion, a connected component will split, and we will have to assign new roots. We define the \emph{stretch} of an edge in a rooted spanning forest.
\begin{definition}[Stretches of {$F$}]
  \label{def:stretchf}
  Given a rooted spanning forest $F$ of a graph $G = (V, E)$ with lengths $\bell \in \R_{>0}^E$, the \emph{stretch} of an edge $e = (u, v) \in E$ is given by
  \begin{align*}
    \str^{F,\bell}_e \defeq
    \begin{cases}
      1 + \left\langle \bell, |\bp(F[u,v])|\right\rangle/\bell_e &~\text{ if } \root^F_u = \root^F_v \\
      1 + \left\langle \bell, |\bp(F[u, \root^F_u])| + |\bp(F[v, \root^F_v])| \right\rangle/\bell_e &~\text{ if } \root^F_u \neq \root^F_v,
    \end{cases}
  \end{align*}
  where $\bp(F[\cdot,\cdot]),$ as defined in \cref{sec:prelim}, maps a path to its signed indicator vector.
\end{definition}
\cite{CKLPPS22} shows how to dynamically maintain a low-stretch decomposition with stretch upper bounds, along with some other useful properties that allow a cleaner interaction with the dynamic spanner that we introduce later in \cref{thm:spanner}.
\begin{lemma}[Dynamic low-stretch decomposition {\cite[Lemma 6.5]{CKLPPS22}}]
  \label{lemma:dynamiclsd}
  Let $G = (V, E)$ be a dynamic graph with at most $m$ edges at all times, with lengths $\bell \in \R^E_{>0}$, weights $\bv \in \R^E_{>0}$, and parameter $k \in \Z_{>0}$. There is a deterministic algorithm with total runtime $\O(m)$ that initializes a tree $T$ spanning $V$, and a rooted spanning forest $F \subseteq T$, a edge-disjoint partition $\cW$ of $F$ into $O(m/k)$ sub trees and stretch overestimates $\wstr_e$.
  The algorithm maintains $F$ (which is decremental) under updates to $G$, such that $\wstr_e \defeq 1$ for any new edge $e$ added by edge insertions, and:
  \begin{enumerate}
  \item $F$ initially has $O(m/k)$ connected components and $O(q \log^2 n)$ more connected components after $q$ total edge updates for any $q \le \O(m)$.  \label{item:comp}
  \item $\str^{F,\bell}_e \le \wstr_e \le O(k \log^6 n)$ for all $e \in E$ at all times, including inserted edges $e$. \label{item:stretch}
  \item $\sum_{e \in E^{(0)}} \bv_e \wstr_e \le O(\|\bv\|_1 \log^4 n)$, where $E^{(0)}$ is the initial edge set of $G$. \label{item:avgstretch}
  \item
  Initially, $\cW$ contains $O(m/k)$ subtrees.
  For any piece $W \in \cW, W \subseteq V$, $|\partial W| \le 1$ and $\vol_G(W \setminus R) \le O(k\log^2n)$ at all times, where $R \supseteq \partial \cW$ is the set of roots in $F$.
  Here, $\partial W$ denotes the set of \emph{boundary vertices} that are in multiple partition pieces.
    \label{item:deg}
  \end{enumerate}
\end{lemma}
It is critical that we are maintaining stretch overestimates which do not change as $F$ undergoes deletions, because we cannot afford to update every single edge whose stretch changes.
By running a multiplicative weight scheme, we can construct an average of $\O(k)$ low-stretch decompositions whose average stretch on every edge is $\O(1)$.
\begin{lemma}[MWU {\cite[Lemma 6.6]{CKLPPS22}}]
  \label{lemma:mwu}
  There is a deterministic algorithm that on a graph $G = (V, E)$ with lengths $\bell$ and a positive integer $k$ computes $t$ spanning trees, rooted spanning forests, and stretch overestimates $\{(T_i, F_i \subseteq T_i, \wstr^i_e)\}_{i=1}^t$ (\cref{lemma:dynamiclsd}) for some $t = \O(k)$ such that
  \begin{align}
    \label{eq:strMWU}
    \sum_{i\in[t]} \blambda_i \wstr^{i}_e \le O(\log^4 n) \forall e \in E,
  \end{align}
  where $\blambda \in \R_{>0}^{[t]}$ is the uniform distribution over the set $[t]$, i.e. $\blambda = \vec{1} / t.$ The algorithm runs in $\O(mk)$-time.
\end{lemma}
Given a low-stretch decomposition corresponding to a forest $F$, we will recursively process the graph $G/F$, i.e. $G$ with the forest edges contracted. We call $G/F$ the \emph{core graph}. In the definition below, we describe how we define the gradients and lengths of the core graph. The goal is to be able to eventually formally argue that routing a flow to the core graph on average preserves its gradient, and only increases its length by $\O(1)$.
\begin{definition}[Core graph]
  \label{def:coregraph}
  Consider a graph $G$ with lengths $\bell \in \R^E_{>0}$, and rooted spanning forest $F \subseteq T$ with stretch overestimates $\wstr_e$ satisfying the guarantees of \cref{lemma:dynamiclsd}. Define the \emph{core graph} $\cC(G, F)$ as a graph with the same edge and vertex set as $G/F$. For $e = (u, v) \in E(G)$ with image $\hat{e} \in E(G/F)$ define its length as $\bell^{\cC(G,F)}_{\hat{e}} \defeq \wstr_e\bell_e$ and gradient as $\bg^{\cC(G,F)}_{\hat{e}} \defeq \bg_e + \l \bg, \bp(T[v, u]) \r$.
\end{definition}
The reason for the tree $T$ is to ensure that even as $F$ undergoes edge deletions that the gradient $\bg^{\cC(G, F)}_{\hat{e}}$ does not change. As noted in \cite[Lemma 7.13]{CKLPPS22}, this definition preserves the gradient of any cycle/circulation. We give a more formal statement of our required statement in \cref{lemma:liftcycle}.

Note that the core graph, which has edge set $G/F$, still has about $m$ edges, but only has at most $\O(m/k)$ vertices if $q \le \O(m/k)$ in \cref{lemma:dynamiclsd}. Towards this, we now define a spanner of the core graph (which we call a \emph{sparsified core graph}). The main guarantees are that it is a spanner with short paths explicitly maintained by the data structure.
\begin{definition}[Spanner with embedding]
\label{def:spanner}
Given a graph $G$ with $m$ edges, $n$ vertices, and parameter $k$, we say that subgraph $H$ is a \emph{$(\gamma_s, \gamma_l)$-spanner} of $G$ with embedding $\Pi_{G\to H}$ if:
\begin{enumerate}
\item $\Pi_{G\to H}(e)$ is a path between the endpoints of $e$ in $H$ with at most $\gamma_l$ edges.
\item For any $e \in E(G)$, all edges $e' \in \Pi_{G\to H}(e)$ satisfy $\bell_e^G \approx_2 \bell_{e'}^G$.
\item $H$ has at most $(m/k+n)\gamma_s$ edges.
\item The lengths and gradients of edges in $H$ are the same as in $G$.
\end{enumerate}
\end{definition}
We can now define a sparsified core graph.
\begin{definition}[Sparsified core graph]
  \label{def:sparsecore}
  Given a graph $G$, forest $F$, and parameter $k$, let $\cS(G, F) \subseteq \cC(G, F)$ and embedding $\Pi_{\cC(G, F) \to \cS(G, F)}$ be a $(\gamma_s, \gamma_l)$-spanner with embedding of $\cC(G, F)$ (\cref{def:spanner}). We say that $\cS(G, F)$ is a $(\gamma_s, \gamma_l)$-\emph{sparsified core graph}.
\end{definition}
Finally, we define the overall structure for our data structure for dynamically maintaining min-ratio cycles against adaptive adversaries. Our data structures are recursive. Our first goal is to reduce the number of edges $m$ to $\Ohat(n)$, at the cost of a small approximation factor. Normally, this would be done by simply maintaining a spanner of the original graph. However, because our data structure requires explicit routings and to succeed against adaptive adversaries, we need to reduce the number of edges by a factor of $k$ each level, instead of by $m/n$ all at once. This way, the total amortized runtime will be $\poly(k)$, which is acceptable.

\begin{definition}[Spanner chain]
\label{def:spannerchain}
For a graph $G$, parameter $k$, and depth $d_s$, we say that a sequence of graphs $\{G_0, G_1, \dots, G_{d_s}\}$ is a $(\gamma_s, \gamma_l)$-\emph{spanner chain} if $G_0 = G$ and $G_{i+1}$ is a $(\gamma_s, \gamma_l)$-spanner of $G_i$ (\cref{def:spanner}) with embedding $\Pi_{G_i\to G_{i+1}}$ for all $i \in \{0, 1, \dots, d_s - 1\}$.
\end{definition}
The number of levels $d_s$ will satisfy $k^{d_s} \approx m/n$.

We now state a variation on the previous definition that allows for vertex reduction, as well as \emph{branching}, i.e. a graph $G_i$ generates several graphs $G_{i+1}$ that are all different sparsified core graphs of $G_i$. Our data structure will have $d_t$ levels, where $k^{d_t} \approx \sqrt{n}$.
\begin{definition}[{$B$}-Branching Tree Chain]
  \label{def:chain}
  For a graph $G$, parameter $k$, and branching factor $B$, a $B$-\emph{branching tree-chain} consists of collections of graphs $\{\cG_i\}_{0 \le i \le d_t}$,  such that $\cG_0 \defeq \{G\}$, and we define $\cG_i$ inductively as follows,
  \begin{enumerate}
  \item For each $G_i \in \cG_i$,  $i < d_t,$  we have a collection of $B$ trees $\cT^{G_i} = \left\{T_1, T_2, \dots, T_B\right\}$ and a collection of $B$ forests $\cF^{G_i} = \left\{F_1, F_2, \dots, F_B \right\}$ such that $E(F_j) \subseteq E(T_j)$ satisfy the conditions of \cref{lemma:dynamiclsd}.
  \item For each $G_i \in \cG_i$, and $F \in \cF^{G_i},$ 
  we maintain $(\gamma_s,\gamma_l)$-sparsified core graphs and embeddings $\cS(G_i, F)$ and $\Pi_{\cC(G_i, F) \to \cS(G_i, F)}$.
  \item We let $\cG_{i+1} \defeq \{\cS(G_i, F) : G_i \in \cG_i, F \in \cF^{G_i}\}$.
  \end{enumerate}

  We let a \emph{tree-chain} be a single sequence of graphs $G_0, G_1, \dots, G_{d_t}$ such that $G_{i+1}$ is the $(\gamma_s,\gamma_l)$-sparsified core graph $\cS(G_i, F_i)$ with embedding $\Pi_{\cC(G_i, F_i) \to \cS(G_i, F_i)}$ for some $F_i \in \cF^{G_i}$ for $0 \le i < d_t$.
\end{definition}
We will ultimately show that if $B = \O(k)$ then we can dynamically maintain a spanner-chain of a graph $G$ of depth $d_s$, and then a $B$-branching tree-chain of depth $d_t$ on the spanner in amortized $\Ohat(\sqrt{n})$ time. Then we will analyze how to query a cycle from this data structure.

We conclude this section by describing how to convert a tree-chain into a forest on a graph, and how a branching tree chain extends to a collection of forests.
\begin{definition}[Forests from tree chain]
\label{def:treechaintrees}
  Given a graph $G$ and tree chain $G_0, G_1, \dots, G_{d_t}$ where $G_0 = G$, define the corresponding forest $F^{G_0, G_1, \dots, G_{d_t}} \defeq \bigcup_{i=0}^{d_t} F_i$ of $G$ as the union of preimages of edges of $F_i$ in $G = G_0$.

  Define the set of trees corresponding to a branching tree-chain of graph $G$ as the union of $F^{G_0, G_1, \dots, G_{d_t}}$ over all tree-chains $G_0, G_1, \dots, G_{d_t}$ where $G_0 = G$: \[ \cF^G \defeq \{ T^{G_0, G_1, \dots, G_{d_t}} : G_0, G_1, \dots, G_{d_t} \text{ s.t. } G_{i+1} = \cS(G_i, F_i) \forall 0 \le i < d_t \}. \]
\end{definition}
Note that the forest $F^{G_0, G_1, \dots, G_{d_t}}$ has exactly $|V(G_{d_t})|$ connected components. Additionally, given a spanning tree $T$ on $G_{d_t}$, adding the preimages of edges in $T$ to $F^{G_0, G_1, \dots, G_{d_t}}$ creates a spanning tree of $G$.

\subsection{Runtime for constructing data structures}
\label{sec:runtimeofds}

Towards formally analyzing the runtimes for maintaining the data structures defined in \cref{sec:dsdef}, we state a result from \cite{CKLPPS22} on dynamically maintaining spanners under vertex splits (and with explicit path embeddings).
\begin{theorem}[Dynamic spanner with vertex splits {\cite[Theorem 5.1]{CKLPPS22}}]
\label{thm:spanner}
There is a dynamic algorithm that given an unweighted dynamic graph $G$ with $m$ edges, $n$ vertices with maximum degree $\Delta$ at all times undergoing edge insertions, deletions, and at most $\O(n)$ vertex splits, maintains a subgraph $H$ of $G$ and an explicit embedding $\Pi_{G\to H}$ satisfying that:
\begin{enumerate}
\item $H$ has at most $n\gamma_s$ edges, and amortized recourse $\gamma_r$.
\item For all $e \in E(G) \setminus E(H)$, the path $\Pi_{G\to H}(e)$ contains $\gamma_l$ edges.
\end{enumerate}
where $\gamma_s,\gamma_r,\gamma_l = \exp(O((\log m)^{3/4} \log \log m))$
The algorithm has initialization time $\Ohat(m)$ and runs in amortized $\Ohat(\Delta)$ time per update.
\end{theorem}
We state a corollary in the case where there are no vertex splits and there is no upper bound on the degrees of $G$.
\begin{corollary}[Dynamic spanner]
\label{cor:spanner}
There is a dynamic algorithm that given a dynamic graph $G$ with $m$ edges, $n$ vertices, and parameter $k$, undergoing edge insertions and deletions maintains a subgraph $H$ of $G$ and an explicit embedding $\Pi_{G\to H}$ satisfying that:
\begin{enumerate}
\item $H$ has at most $\gamma_s(m/k + n)$ edges, and amortized recourse $\gamma_r$.
\item For all $e \in E(G) \setminus E(H)$, the path $\Pi_{G\to H}(e)$ contains $\gamma_l$ edges,
\end{enumerate}
where $\gamma_s, \gamma_r, \gamma_l \le \exp(O((\log m)^{3/4} \log \log m))$.
The algorithm has initialization time $\Ohat(m)$ and runs in amortized $\Ohat(k)$ time per update.
\end{corollary}
\begin{proof}
	Given $G$, create a new graph $G'$ as follows: Repeatedly split each high degree vertex until the degrees are at most $k-1$.
	The graph $G'$ now has at most $O(n+m/k)$ vertices.
	
	If an edge is inserted to a vertex $v$ in $G'$ that already has degree $k-1$, then instead create a new vertex $v'$ and connect the edge with $v'$. All future insertion to $v$ will be inserted into $v'$ instead.
	
	To obtain \Cref{cor:spanner} we now run \Cref{thm:spanner} on $G'$.
	Let $H'$ be the spanner obtained on $G'$.
	Note that after contracting the vertices representing $v\in V(G)$ in $H'$, we obtain a spanner in $G$ because the maximum length of paths only decreases.
	
	Note that \Cref{thm:spanner} does not support new vertex insertions.
	The creation of a new vertex in $G'$ can be handled by adding an additional $O(m/k)$ isolated vertices during initialization. Instead of creating a new vertex during an edge insertion, we use one of these isolated vertices instead.
	After $O(m)$ edge insertion there may be no isolated vertices left, in which case restart the algorithm. Given the almost-linear initialization complexity, the complexity amortizes. Further, the amortized recourse increases by only an additive $\gamma_s O((m/k+n)/m) = O(\gamma_s)$.
\end{proof}
These results allow us to build a chain of spanners to reduce the number of edges in the original graph $G$ to $\Ohat(n)$, with recourse $\Ohat(1)$.
\begin{lemma}[Dynamic spanner chain]
\label{lemma:dynamicspannerchain}
There is an algorithm that given size reduction parameter $k$, depth $d_s$, and a graph $G$ undergoing edge insertions and deletions, maintains $(\gamma_s,\gamma_l)$-spanner chain (see \Cref{def:spannerchain}) $\{G_0, \dots, G_{d_s}\}$ in amortized $\Ohat(k\gamma_r^{d_s})$ time, and $G_{d_s}$ has amortized recourse $O(\gamma_r^{d_s})$.
Here $\gamma_l,\gamma_r \le \exp(O((\log m)^{3/4} \log \log m))$ and $\gamma_s \le \exp(O((\log m)^{3/4} \log \log m))\cdot\log W$ where $W$ is the ratio of largest to smallest length $\bell\in\R_{>0}^E$ in $G$.
\end{lemma}
\begin{proof}
To maintain a $(\gamma_s,\gamma_l)$-spanner of any graph $G$, we split the graph into graphs $G_i$ where each $G_i$ contains the edges with $\bell_e \in (2^i, 2^i+1]$.
So we have $O(\log W)$ such graphs.
Then we run \Cref{cor:spanner} on each $G_i$ and take the union $H$ of the maintained $H_i$.
This gives a $(\gamma_s,\gamma_l)$-spanner with $\gamma_r$ recourse for
$\gamma_l,\gamma_r \le \exp(O((\log m)^{3/4} \log \log m))$ and $\gamma_s \le \exp(O((\log m)^{3/4} \log \log m))\cdot\log W$.
Note that property 1 and 3 of \Cref{def:spanner} are satisfied by \Cref{cor:spanner}.
Property 4 is satisfied by copying the length and gradient values from $G_i$ to $H_i$.
Property 2 is satisfied since each embedding only yields a path in a single $G_i$, in which all edge weights are of the same length up to a factor $2$.

Now consider the task of maintaining a spanner chain for some input graph $G$.
To solve this, define $G_0 \defeq G$ and run the above algorithm on $G_0$. Let $H$ be the output and define $G_1 \defeq H$. Repeat this recursively for $d_s$ levels, i.e.~$G_{i+1}$ is the output of running above algorithm on $G_i$ for $i < d_s$.
This results in a $(\gamma_s,\gamma_l)$-spanner chain.

The recourse is bounded by $\gamma_r^{d_s}$ since one update to $G_i$ results in $\gamma_r$ changes to $G_{i+1}$ and we have $d_s$ levels.
\end{proof}

Now we turn to building the recursive data structure. Towards this, it is useful to define how to route a circulation down from a graph $G$ to the core graph $\cC(G, F)$, and to the sparsified core graph $\cS(G, F)$. To verify that our recursive data structure succeeds, the critical piece will be to bound the length increase under these routings.
\begin{definition}[Routing flow into core graph]
\label{def:routeflowcore}
For graph $G$, forest $F \subseteq T$, and circulation $\bc^G \in \R^{E(G)}$, we define the circulation $\bc^{\cC(G, F)} \in \R^{E(\cC(G, F))}$ as
$\bc^{\cC(G, F)}_{\hat{e}} = \bc^G_e$ where $\hat{e}$ is the projection of $e$ into $\cC(G, F)$.
\end{definition}

\begin{definition}[Routing flow into spanner]
\label{def:routeflowspanner}
Given a spanner $(\gamma_s,\gamma_l)$-spanner $H$ of $G$ with embedding $\Pi_{G\to H}$, and circulation $\bc^G \in \R^{E(G)}$, define the circulation $\bc^H \in \R^{E(H)}$ as $\bc^H \defeq \sum_{e \in E(G)} \bc^G_e \Pi_{G\to H}(e)$, where $\Pi_{G\to H}(e) \in \R^{E(H)}$ is viewed as a flow vector.
\end{definition}

Combining \cref{def:routeflowcore,def:routeflowspanner} allows us to start from a circulation $\bc^G \in \R^{E(G)}$, and a tree-chain $\{G_0, \dots, G_{d_t}\}$ (\cref{def:chain}) and recursively define circulations $\bc^{G_i} \in \R^{E(G_i)}$ for all $0 \le i \le d_t$, because $G_{i+1}$ is a spanner of $\cC(G_i, F_i)$ for some forest $F_i \subseteq E(G_i)$.

Finally, we state the construction of our full recursive data structure: the $\O(k)$-branching tree chain.
\begin{lemma}[Dynamic branching tree-chain]
\label{lemma:dynamicchain}
There is an algorithm that given a graph $G$ undergoing edge insertions and deletions, and parameter $k$ and depth $d_t$ maintains a $\O(k)$-branching tree chain $\{\cG_0, \dots, \cG_{d_t}\}$, and a distribution $\blambda$ over $\cG_{d_t}$ such that
\begin{align} \Pr_{G_{d_t} \sim \blambda}\left[\|\mL^{\cC(G_i, F_i)} \bc^{\cC(G_i, F_i)} \|_1 \le \O(1) \cdot \|\mL^{G_i} \bc^{G_i} \|_1 \forall i \in \{0, 1, \dots, d_t\} \right] \ge 1/2, \label{eq:dsstretch} \end{align}
where $\{G_0, \dots, G_{d_t}\}$, where $G_{i+1} = \cS(G_i, F_i)$ for forests $F_0, \dots, F_{d_t-1}$ is the tree chain ending at $G_{d_t}$. The amortized runtime is $\poly(k) \cdot \O(k\gamma_r)^{d_t}$.
\end{lemma}
Towards proving \cref{lemma:dynamicchain}, specifically \eqref{eq:dsstretch}, we will prove that to the average of core graphs constructed via \cref{lemma:mwu} only increases the total length on average by a $\O(1)$ factor.
\begin{lemma}[Average tree stretch]
\label{lemma:avgtreestretch}
Let $G$ be a graph, and let $F_1, \dots, F_t$ for $t = \O(k)$ be constructed to satisfy the conclusion of \cref{lemma:mwu}, and $\blambda$ is uniform on $[t]$. Then for all flows $\bc^G \in \R^{E(G)}$,
\[ \E_{i\sim\blambda}\left[\|\mL^{\cC(G, F_i)} \bc^{\cC(G, F_i)}\|_1 \right] \le \O(1) \cdot \|\mL^G \bc^G\|_1. \]
\end{lemma}
\begin{proof}
Recall that $\bc_{\hat{e}}^{\cC(G, F_i)} = \bc^G_e$ (\cref{def:routeflowcore}) and $\bell^{\cC(G,F_i)}_{\hat{e}} \defeq \wstr_e^i\bell_e$ (\cref{def:coregraph}). Thus
\[ \E_{i\sim\blambda}\left[\bell^{\cC(G,F_i)}_{\hat{e}} |\bc^{\cC(G,F_i)}_{\hat{e}}| \right] = |\bc^G_e| \sum_{i \in [t]} \blambda_i \wstr_e^i \le O(\log^4 m) \cdot |\bc^G_e|, \] by \cref{lemma:mwu}. The result follows by linearity of expectation.
\end{proof}
We can now show \cref{lemma:dynamicchain} by combining our dynamic low-stretch decomposition data structure (\cref{lemma:dynamiclsd}), dynamic spanner (\cref{thm:spanner}), and \cref{lemma:avgtreestretch}. The proof closely follows the analogous result in \cite[Lemma 7.8]{CKLPPS22}.
\begin{proof}[Proof of \cref{lemma:dynamicchain}]
We describe how to recursively maintain a $\O(k)$-branching tree chain. For a graph $G_i$, maintain forests $F_1, \dots, F_t \subseteq E(G_i)$ for $t = \O(k)$ satisfying the guarantees of \cref{lemma:mwu} with distribution $\blambda^{G_i}$, by using the dynamic low-stretch decomposition data structure in \cref{lemma:dynamiclsd}. As $F_j$ changes decrementally via \cref{lemma:dynamiclsd}, the core graph $\cC(G_i, F_j)$ undergoes updates, for which we use the dynamic spanner in \cref{thm:spanner} to maintain the sparsified core graph $\cS(G_i, F_j)$. We do this for $d_t$ layers. When a graph $G_i$ receives $E(G_i)/k$ total updates, we rebuild the forests $F_1, \dots, F_j$.

To verify \cref{lemma:dynamicchain}, we must check the following pieces. First, we need to check that the graphs $\cC(G_i, F_j)$ have bounded maximum degree, and undergo edge insertions/deletions, and vertex splits, so that \cref{thm:spanner} is applicable. We then check \eqref{eq:dsstretch}, and finally bound the runtime.

\paragraph{Applying the dynamic spanner.} Recall that by \cref{lemma:dynamiclsd}, each forest $F_j$ is decremental, and $G_i$ undergoes edge insertions/deletions. Additionally, by the definitions of $\wstr_e^j$ and $\bg^{\cC(G_i, F_j)}$ in \cref{def:coregraph}, edges $e \in G_i$ keep the same length and gradient even under deletions to $F_j$. Hence the graph $\cC(G_i, F_j)$ undergoes edge insertions/deletions, and vertex splits.

Using $\cC(G_i, F_j)$, we build a modified graph $\cC^j$ in to bound the maximum degree by $\O(k)$. Let $R$ be the initial set of roots in $F_j$. For each root $r \in R$, we split off a vertex $u_r$ that is adjacent to all edges in $G_i$ adjacent to $r$. Note that the vertices $u_r$ will never be split further as the result of edge deletions to $F_j$, because the set of roots is incremental. Hence, as in the proof of \cref{cor:spanner}, we can further initially further split each vertex $u_r$ into vertices of degree at most $\O(k)$. For all partition pieces $W \in \mathcal{W}$, we split off a vertex $u_W$ adjacent to vertices in $W \setminus R$. By item \ref{item:deg} of \cref{lemma:dynamiclsd}, we know $\deg(u_W) \le \vol_G(W \setminus R) \le \O(k)$. Let the result of splitting off these vertices $u_r$ and $u_W$ be $\cC^j$. Because $|R| \le O(m/k)$ and $|\mathcal{W}| \le O(m/k)$ initially, $\cC^j$ still has $O(m/k)$ vertices. Hence we may apply \cref{thm:spanner} with $\Delta = \O(k)$. Note that by contracting vertices, a $(\gamma_s,\gamma_l)$-spanner on $\cC^j$ with embedding corresponds to one on $\cC(G_i, F_j)$.

\paragraph{Bounding the stretch.} Let the distribution $\blambda$ on $\cG_{d_t}$ be defined as follows: starting from $G = G_0$, sample $G_{i+1} \sim \blambda^{G_i}$ for $i = 0, 1, \dots, d_t - 1$. By \cref{lemma:avgtreestretch} and Markov's inequality, \[ \Pr_{G_{i+1} \sim \blambda^{G_i}}\left[ \|\mL^{\cC(G_i, F_i)} \bc^{\cC(G_i, F_i)}\|_1 \le \O(1) \cdot \|\mL^{G_i} \bc^{G_i}\|_1 \right] \ge 1 - \frac{1}{2 \log m}, \] by increasing the $\O(1)$ by a $2 \log m$ factor. Thus \eqref{eq:dsstretch} follows by a union bound over the $d_t \le \log m$ layers of the data structure.

\paragraph{Amortized runtime.} Per update to a graph $G_i \in \cG_i$, a single graph $\cC(G_i, F_j)$ undergoes $\O(1)$ edge insertions/deletions, and vertex splits by \cref{lemma:dynamiclsd}. Thus, in total the graphs $\cC(G_i, F_j)$ for $j = 1, \dots, t$ for $t \le \O(k)$ undergo $\O(k)$ such operations. Using the spanner in \cref{thm:spanner} to maintain each of $\cS(G_i, F_j)$ incurs additional recourse $O(\gamma_r)$ by \cref{thm:spanner}. Thus, the recourse per layer is $\O(k\gamma_r)$. Because $\Delta = \O(k)$, and applying the runtimes in \cref{lemma:mwu,lemma:dynamiclsd}, the additional runtime has $\poly(k)$ overhead. Thus the total runtime is $\poly(k) \cdot \O(k\gamma_r)^{d_t}$.
\end{proof}

\subsection{Querying a min-ratio cycle}
\label{sec:query}

The goal of this section is to describe how to go from a spanner chain and branching tree chain to actually recovering a circulation $\bDelta$ with small ratio. We start by analyzing how to find cycles in a graph $G$ given a spanner $H$ and embedding $\Pi_{G\to H}$.
\begin{definition}[Sparsifier cycles]
\label{def:sparsifiercycle}
For a graph $G$ and spanner $H$ with embedding $\Pi_{G\to H}$, for edge $e \in E(G) \setminus E(H)$ we define the \emph{sparsifier cycle} of $e$ to be the circulation $\ba(e) \in \R^{E(G)}$ representing the cycle $e \oplus \Pi_{G\to H}(e)$.
\end{definition}
We observe that either $H$ supports a good min-ratio cycle, or some spanner cycle has good ratio. Because our algorithm can explicitly maintain the embedding, we can efficiently determine whether we have found a good enough cycle, or whether we should recursively try to find a cycle in $H$.
\begin{lemma}[Spanner cycle or recurse]
\label{lemma:dichotomy}
For a graph $G$ with gradient $\bg \in \R^{E(G)}$ and lengths $\bell \in \R^{E(G)}_{\ge0}$, define
\[ \OPT = \min_{\substack{\mB^\top\bc=0 \\ \bc\in\R^{E(G)}}}  \frac{\bg^\top \bc}{\|\mL \bc\|_1}. \] For $(\gamma_s,\gamma_l)$-spanner $H$ with embedding $\Pi_{G\to H}$, we have
\[ \min_{e \in E(G) \setminus E(H)} \frac{\bg^\top \ba(e)}{\|\mL \ba(e)\|_1} \le \frac{\OPT}{5\gamma_l} \enspace \enspace or \enspace \enspace \min_{\substack{\mB^\top\bc=0 \\ \bc\in\R^{E(H)}}} \frac{\bg^\top \bc}{\|\mL \bc\|_1} \le \frac{\OPT}{5\gamma_l}. \]
\end{lemma}
\begin{proof}
Let $\bc^*$ be an optimal solution to
$$
\min_{\substack{\mB^\top\bc=0 \\ \bc\in\R^{E(G)}}}  \frac{\bg^\top \bc}{\|\mL \bc\|_1}.
$$
We can map this circulation onto $H$ by mapping the flow $\bc^*_e$ of each $e\in E(G)\setminus E(H)$ onto the path $\Pi_{G\to H}(e)$.
This way we obtain a circulation $\bc'$ on $H$ defined as
$$
\bc' = \bc^* - \sum_{e\in E(G)\setminus E(H)} \bc^*_e \ba(e)
$$
and thus
$$
|\bg^\top \bc^* |
\le
|\bg^\top \bc'| + \sum_{e\in E(G)\setminus E(H)} |\bc^*_e\bg^\top \ba(e)|.
$$

Note that all edges in $\ba(e)$ have the same length up to a factor of $2$, so $\|\bc^*_e \mL \ba(e)\|_1 \le 2\bell_e|\bc^*_e|\gamma_l$.
This implies
$$
\|\mL \bc'\|_1 + \sum_{e\in E(G)\setminus E(H)} |\bc^*_e| \|\mL \ba(e)\|_1 
\le
\|\mL \bc^*\|_1 + 2\sum_{e\in E(G)\setminus E(H)} |\bc^*_e| \|\mL \ba(e)\|_1 
\le
5\gamma_l\|\mL\bc^*\|_1,
$$

In summary,
\begin{align*}
\frac{|\OPT|}{5 \gamma_l}
=&~
\frac{|\bg^\top \bc^*|}{\|\mL \bc^*\|_1} \frac{1}{5 \gamma_l}
\le
\frac
{ |\bg^\top \bc'| + \sum_{e\in E(G)\setminus E(H)} |\bc^*_e \bg^\top \ba(e)| }
{ \|\mL \bc'\|_1 + \sum_{e\in E(G)\setminus E(H)} |\bc^*_e| \|\mL \ba(e)\|_1 } \\
\le&~ 
\max\left\{ 
\frac{|\bg^\top \bc'|}{\|\mL \bc'\|_1}
,
\max_{e\in E(G)\setminus E(H)}
\frac{|\bc^*_e|\cdot|\bg^\top \ba(e)|}{|\bc^*_e| \|\mL \ba(e)\|_1}
\right\}
\end{align*}
Canceling the $|\bc^*_e|$ concludes \Cref{lemma:dichotomy}.
\end{proof}
The following lemma says that finding cycles in ``lower levels'' of the tree-chain directly gives cycles in the original graph $G$ that can be represented by $m^{o(1)}$ paths on a tree.
\begin{lemma}[Lifted cycle {\cite[Lemma 7.13, Lemma 7.14]{CKLPPS22}}]
\label{lemma:liftcycle}
Let $G_0, \dots, G_{d_t}$ be a tree chain, and $\hat{C} = \hat{e_1} \oplus \hat{e_2} \oplus \dots \oplus \hat{e_L}$ is a cycle in $\cC(G_i, F_i)$, representing by the circulation vector $\hat{\bc} \in \R^{E(\cC(G_i, F_i))}$. Then there is a cycle $C$ in $G$ with circulation vector $\bc \in \R^{E(G)}$ satisfying
\[ \frac{\langle \bg^G, \bc\rangle}{\|\mL^G \bc\|_1} \le \frac{\langle \bg^{\cC(G_i, F_i)}, \hat{\bc} \rangle}{\|\mL^{\cC(G_i, F_i)} \hat{\bc}\|_1}. \]
Additionally, $C$ consists of the preimage of edges $\hat{e_1}, \dots, \hat{e_L}$ in $G$ (which we denote as $e_j = (u_j, v_j)$), along with the paths connecting $v_j$ to $u_{j+1}$ for $1 \le j \le L$ (where $u_{L+1} = u_1$) in $F^{G_0,\dots,G_{d_t}}$.
\end{lemma}
Also, \cite[Lemma 7.13]{CKLPPS22} shows that a cycle $\bc$ and its mapping $\bc^{\cC(G, F)}$ into the core graph have the same gradient.
\begin{lemma}[\!\!{\cite[Lemma 7.13]{CKLPPS22}}]
\label{lemma:samegradient}
For a graph $G$ with forest $F \subseteq E(G)$, and circulation $\bc^G \in \R^{E(G)}$ we have $\l \bg^G, \bc^G \r = \l \bg^{\cC(G, F)}, \bc^{\cC(G, F)} \r$.
\end{lemma}
We can now combine all these pieces to show \cref{thm:minratio}.
\begin{proof}[Proof of \cref{thm:minratio}]
Set $k = \exp(O(\log^{7/8} m))$.
Let $\mathcal{D}^{(SC)}$ be a data structure for maintaining a spanner chain (\cref{lemma:dynamicspannerchain}), and let $\mathcal{D}^{(BTC)}$ be a data structure that maintains a branching tree chain as in \cref{lemma:dynamicchain}, with the parameter choices $d_s, d_t$ be maximal, and satisfy $k^{d_s} \le m/n$ and $k^{d_t} \le \sqrt{n}$, so that $d_s, d_t \le \log^{1/8} m$.

Given a dynamic graph $G$, we first pass $G$ to $\mathcal{D}^{(SC)}$. This data structure maintains a spanner chain with graphs $G_0, \dots, G_{d_s}$ -- let the bottom level graph in this chain be $H \defeq G_{d_s}$. By the choice of $d_s$, $H$ will have $\Ohat(n)$ edges, and also undergo $\Ohat(1)$ edge updates per single update to $G$. We pass these updates to $\mathcal{D}^{(BTC)}$, maintained starting at $H$. Let the collections graphs in the $\O(k)$-branching tree chain be $\mathcal{H}_0, \dots, \mathcal{H}_{d_t}$, where $\mathcal{H}_i$ contains all graphs on level $i$. Because the branching factor is $\O(k)$, we can maintain the $\O(k)^{d_t} \le \Ohat(\sqrt{n})$ forests given as in definition \cref{def:treechaintrees}. We now describe how to implement edge insertions/deletions.

\paragraph{Edge insertions and deletions.} Maintain the data structure exactly as described in the above paragraph. By \cref{lemma:dynamicspannerchain}, $H$ has amortized recourse $\gamma_s^{d_s} \le \exp(O(\log^{7/8} m \log \log m))$. These updates are passed to $\mathcal{D}^{(BTC)}$, which runs in amortized time \[ \poly(k) \cdot \O(k\gamma_r)^{d_t} \le \sqrt{n} \exp(O(\log^{7/8} m \log \log m)) \] per update to $H$.

\paragraph{Query.} For a graph $G'$, let $\OPT^{G'} \defeq \min_{\mB^\top\bc=0} \frac{\l \bg^{G'}, \bc \r}{\|\mL^{G'} \bc\|_1}$. Our goal is to find a cycle whose ratio is at most $\kappa \OPT^G$. The implementation of \textsc{Query} has three cases.
\begin{itemize}
\item In the first case, there is some $i \in \{0, 1, \dots, d_s-1\}$ and $e \in E(G_i) \setminus E(G_{i+1})$ with \[ \l \bg^{G_i}, \ba(e) \r/\|\mL^{G_i} \ba(e)\|_1 \le \kappa \OPT^G. \] We can then simply return $\ba(e)$, which is a cycle in $G$. Otherwise, note that $\OPT^{G_{d_s}} \ge (5\gamma_l)^{-d_s} \OPT^G$ by induction and \cref{lemma:dichotomy}.
\item In the second case, there is some $i \in \{0, 1, \dots, d_t-1\}$, graph $H_i \in \mathcal{H}_i$, forest $F_i$, and $e \in \cC(H_i, F_i) \setminus \cS(H_i, F_i)$ with $\l \bg^{\cC(H_i, F_i)}, \ba(e) \r/\|\mL^{\cC(H_i, F_i)} \ba(e)\|_1 \le \kappa \OPT^G$. In this case, we can return the lift of $\ba(e)$ to $G$, as defined in \cref{lemma:liftcycle}. Because $\ba(e)$ has at most $\gamma_l$ edges, the lift in $G$ consists of $\gamma_l$ off-tree edges and tree paths by \cref{lemma:liftcycle}.
\item In the third case, we assume neither of the above two cases occurred. In this case, we sample $B \ge O(\log m)$ graphs $H_{d_t} \sim \blambda$ and compute a cycle with value approximately $\OPT^{H_{d_t}}$ statically whp. We claim that if \eqref{eq:dsstretch} holds, then $\OPT^{H_{d_t}} \le \kappa \OPT^G$, so the quality of the cycle returned is good enough with probability $1 - 2^{-B} \ge 1 - m^{-10}$, as desired.

To see this, note that $\OPT^H \le (5\gamma_l)^{-d_s} \OPT^G$ if the first case does not occur. If the second case does not occur, then we deduce that
\[ \OPT^{H_{i+1}} = \OPT^{\cS(H_i, F_i)} \le (5\gamma_l)^{-1} \OPT^{\cC(H_i, F_i)} \le \O(\gamma_l)^{-1} \OPT^{H_i}. \] Here, the last inequality follows because if $\bc^{H_i}$ is the optimal cycle on $H_i$ then
\[ \frac{\l \bg^{\cC(H_i, F_i)}, \bc^{\cC(H_i, F_i)} \r}{\|\mL^{\cC(H_i, F_i)} \bc^{\cC(H_i, F_i)}\|_1} \le \frac{1}{\O(1)} \frac{\l \bg^{H_i}, \bc^{H_i} \r}{\|\mL^{H_i} \bc^{H_i}\|_1}, \] by the guarantee in \eqref{eq:dsstretch} and \cref{lemma:samegradient}. Thus, $\OPT^{H_{d_t}} \le \O(\gamma_l)^{-d_t} \OPT^H \le \kappa \OPT^G$ as desired.
\end{itemize}

\paragraph{Runtime of query.} Cases one and two above can be handled by explicitly maintained the quality of all the cycles $\ba(e)$ at all times, and using a priority queue to maintain the best. This does not add extra runtime beyond simply maintaining the branching tree chain, which costs time \[ \gamma_r^{d_s} \poly(k) \O(k\gamma_r)^{d_t} \le \Ohat(\sqrt{n}). \] In case three, the graph $H_{d_t}$ has at most $\O(\gamma_s/k)^{d_t} O(\gamma_s/k)^{d_s} m \le \Ohat(\sqrt{n})$ edges, so statically computing a cycle requires $\Ohat(\sqrt{n})$ time. To return this cycle, we can add the cycle edges to the corresponding forest from the chain ending at $H_{d_t}$, and reporting the cycle as a single off-tree edge and path. We can then delete the edges we added to revert back to the original forest. This only increases the number of operations by $\Ohat(\sqrt{n})$.
\end{proof}

\subsection{Dynamically updating gradients and lengths}
\label{sec:gl}
The goal of this section is to put everything together to show \cref{thm:inc}. The main missing pieces are to discuss how to maintain the flow implicitly, and how to decide when to update gradients and lengths of edges (to pass to the data structure).

We start by describing what a dynamic link-cut tree can do.
\begin{lemma}[\!\!{\cite[Lemma 3.3]{CKLPPS22}}]
  \label{lem:flowsum}
There is a deterministic data structure that maintains a dynamic tree $T\subset G=(V,E)$ under insertion/deletion of edges with gradient $\bg$ and length $\bell$, and supports the following operations:
\begin{itemize}
  \item Insert/delete edges $e$ to $T$, under the condition that $T$ is always a tree, or update the gradient $\bg_e$ or lengths $\bell_e$.
  \item For a path of vector $\bDelta = \bp(T[u,v])$ for some $u,v\in V$, return $\langle\bg,\bDelta\rangle$ and $\langle\bell,|\bDelta|\rangle$.
  \item Maintain a flow $\bf\in\R^E$ under operations $\bf\leftarrow\bf+\eta\bDelta$ for $\eta\in\R$ and path vector $\bDelta=\bp(T[u,v])$, or query the value $\bf_e$.
\end{itemize} 
\end{lemma}

We also require some ``stability bounds'' on the gradients and lengths to show that the gradients $\wt{\bg}$ and $\wt{\bell}$ do not change more than $\Ohat(m\kappa^{-2}\alpha^{-2})$ times during the method.

The following lemmas are \cite[Lemmas 4.8-4.10]{CKLPPS22}.
\begin{lemma}[Residual stability]
\label{lemma:residualstability}
Let $\wt{\bg} \in \R^E$ satisfy $\left\|\mL_G(\bf)^{-1}\left(\wt{\bg} - \bg_G(\bf)\right) \right\|_\infty \le \eps$ for some $\eps \in (0,1/2]$, and $\wt{\bell} \in \R^E_{>0}$ satisfy $\wt{\bell} \approx_2 \bell_G(\bf)$.
Let $\bDelta$ satisfy $\mB^\top\bDelta = 0$ and $\nicefrac{\wt{\bg}^\top \bDelta}{\left\|\wt{\mL}\bDelta\right\|_1} \le -\kappa'$ for $\kappa' \in (0, 1)$. Then
\[ \frac{|\bc^\top \bDelta|}{\bc^\top \bf - F} \le |\wt{\bg}^\top\bDelta|/(\kappa' m). \]
\end{lemma}

\begin{lemma}[Length stability]
\label{lemma:lengthstability}
If $\|\mL_G(\bf)(\bf-\bar{\bf})\|_\infty \le \eps$ for some $\eps \le 1/100$ then $\bell_G(\bf) \approx_{1+3\eps} \bell_G(\bar{\bf})$.
\end{lemma}

\begin{lemma}[Gradient stability]
\label{lemma:gradientstability}
If $\|\mL_G(\bf)(\bf-\bar{\bf})\|_\infty \le \eps$ and $r \approx_{1+\eps} \bc^\top\bar{\bf} - F$ then $\wt{\bg}$ defined as
\begin{align} \wt{\bg}_e = r/(\bc^\top\bar{\bf}-F) \cdot (20m\bc_e/r + \alpha(\bu_e - \bf_e)^{-1-\alpha} - \alpha(\bf_e + \delta)^{-1-\alpha}) \forall e \in E \label{eq:setbg} \end{align} satisfies
\[ \left\|\mL_G(\bf)^{-1}\left(\wt{\bg} - \bg_G(\bar{\bf})\right) \right\| \le 10\alpha\eps.\]
\end{lemma}

\begin{proof}[Proof of \Cref{thm:inc}]
  We run the IPM \Cref{alg:incremental:ipm} using the data structures of \Cref{thm:minratio,lem:flowsum}.
  For this we let $\kappa = \exp(-O(\log^{7/8}m\log\log m))$ (the parameter from \Cref{thm:minratio}). 
  
  We start with $\bf^{(t)} = 0$ and initialize \Cref{thm:minratio}, $\wt\bell = \bell_G(\bf)$, $\wt\bg = \bg_G(\bf)$.
  This gives us $s = \Ohat(\sqrt{n})$ trees $F_1,\dots,F_s$ on which we run the data structure of \Cref{lem:flowsum}.
  
  Assume for now that we always have $\wt\bg,\wt\bell$ with $\|\mL(\bf^{(t)})^{-1}(\wt{\bg}-\bg_G(\bf^{(t)}))\|_\infty \le \kappa\alpha/32$ and $\wt\bell \approx_2\bell_G(\bf^{(t)})$.
  This is true right after initialization and we will later argue that this will also holds true throughout the algorithm.
  
  \paragraph{Finding $\bDelta$.}
  By \Cref{alg:incremental:ipm} we must attempt to find a circulation $\bDelta\in\R^E$ with $\mB^\top\bDelta=0$ and $\wt\bg^\top\bDelta/\|\wt\mL\bDelta\|_1 \le - \kappa\alpha/4$.
  We attempt to construct such a circle via the \textsc{Query} operation by \Cref{thm:minratio}.
  \Cref{thm:minratio} returns a cycle $\bDelta$ with 
  $$
  \frac{\wt{\bg}^\top\bDelta}{\|\wt\mL\bDelta\|_1} \le \kappa \min_{\mB^\top \bc = 0} \frac{\wt{\bg}^\top\bc}{\|\wt\mL\bc\|_1}.
  $$
  We can compute the left-hand side using the data structures \Cref{lem:flowsum} as \Cref{thm:minratio} provides the cycle $\bDelta$ implicitly as paths on trees.
  If the value is at most $-\kappa\alpha/4$, then we found the desired circulation.
  If the value is larger, then we know
  $$
  \min_{\mB^\top \bc = 0} \frac{\wt{\bg}^\top\bc}{\|\wt\mL\bc\|_1} \ge -\alpha/4
  $$
  so we can pause the IPM and tell the adversary that currently no circulation of cost at most $F$ exists in $G$.
  
  \paragraph{Augmenting $\bf^{(t)}$.}
  If we found an appropriate cycle $\bDelta$ we update the implicit representation of $\bf^{(t)}$ via \Cref{lem:flowsum}.
  For this we directly add $\eta\bDelta_e$ to $\bf^{(t)}_e$ for non-tree edges returned by \Cref{thm:minratio}.
  The edges inside trees have their flow implicitly maintained by \Cref{lem:flowsum}.
  
  \paragraph{Handling edge insertions.}
  When an edge $e$ is inserted by the adversary, we set $\bf^{(t)}_e = 0$, $\wt\bell_e = \bell_G(e)$,
  and insert the edge into \Cref{thm:minratio}.
  
  \paragraph{Maintaining $\wt\bg,\wt\bell$.}
  To maintain $\wt\bg,\wt\bell$ we maintain some approximate flow $\wt\bf$.
  Whenever an edge $e$ is removed from the trees $F_1,\dots,F_s$, we first query the value of the flow on that edge (in the tree) via \Cref{lem:flowsum} before removing the edge using \Cref{lem:flowsum}. We add this to our stored values of $\wt{\bf}_e$. Thus, $\wt{\bf}_e = \bf_e$ for all edges $e$ that are in none of the trees $F_1, \dots, F_s$.
  For the non-tree edges $e$ in cycle $\bDelta$ we also directly update $\wt\bf_e \leftarrow \wt\bf_e + \eta$ so for non tree edges we always have $\wt\bf_e = \bf^{(t)}_e$.
  For the tree edges $e$ that are in cycle $\bDelta$, we will sample edges $\wt\bell_e \leftarrow \bf^{(t)}_e$ with probability proportional to $\eta\wt\bell_e \le 1$, and repeat this $\Ohat(1)$ times.
  By the sampling scheme, whp. we have $|\wt\mL(\bf_e - \wt\bf_e)| < m^{-o(1)}$ on tree edges.
  This sampling can be done efficiently because \cref{lem:flowsum} can maintain sums of $\bell_e$ along paths in a tree.
  
  We now maintain $\wt\bell$ by updating $\wt\bell_e \leftarrow \bell_G(\wt\bf)_e$ whenever $\wt\bf_e$ is updated.
  By \Cref{lemma:lengthstability} we have $\wt\bell\approx_2\bell_G(\bf^{(t)})$.
  
  We also maintain $r \approx_{1+1/(2^{10}\alpha\kappa)} \bc^\top\wt\bf - F$ by updating $r$ whenever this approximation is violated.
  
  We update $\wt\bg_e$ to the expression in \eqref{eq:setbg} for $\bar{\bf} \assign \wt\bf_e$ whenever $\wt\bf_e$ is updated and update all entries of $\wt\bg$ whenever $r$ is changed.
  By \Cref{lemma:gradientstability} we have $\|\mL(\bf^{(t)})^{-1}(\bg-\bg_G(\bf^{(t)}))\|_\infty \le \kappa/8$.
  
  \paragraph{Complexity.}
  We pay $\Ohat(\sqrt{n})$ whenever an entry of $\wt\bg$ or $\wt\bell$ is changed to update the data structures \Cref{thm:minratio} and \Cref{lem:flowsum}.
  By \Cref{lemma:residualstability} we update those values for all edges because of a change in $r$ at most $\Ohat(1)$ times over all iterations of the IPM.
  Additionally, we also perform such an update on an edge $e$ when $\wt\bf_e$ is changed.
  This happens for just $\Ohat(1)$ non-tree edges per iteration as we only have $\Ohat(1)$ non-tree edges in $\bDelta$ by \Cref{thm:spanner}.
  For the tree edges, note that we update an edge with probability $\eta\wt\bell_e$ where $\eta = \Ohat(1/|\wt{\bg}^\top\bDelta|) = \Ohat(1/\|\wt\mL\bDelta\|_1)$. Thus the sampling scheme is valid, and a total of $\Ohat(1)$ edges is updated per iteration through sampling.
\end{proof}

\section*{Acknowledgments}
 We thank Thatchaphol Saranurak for several useful discussions, in particular for introducing us to the problem of thresholded mincost matching and maxflow. We thank Maximilian Probst Gutenberg for pointing out the connection between incremental mincost flow and incremental cycle detection.
Yang P. Liu is supported by the Google PhD Fellowship Program. Aaron Sidford is supported by a Microsoft Research Faculty Fellowship, NSF CAREER Award CCF-1844855, NSF Grant CCF-1955039, a PayPal research award, and a Sloan Research Fellowship.
Part of this work was done while Jan van den Brand was at the Simons Institute for the Theory of Computing and the Max Planck Institute for Informatics. 

{\small
\bibliographystyle{alpha}
\bibliography{refs}}

\newcommand{\etalchar}[1]{$^{#1}$}
\begin{thebibliography}{LMSW22}

\bibitem[ABKL22]{AssadiBKL22}
Sepehr Assadi, Soheil Behnezhad, Sanjeev Khanna, and Huan Li.
\newblock On regularity lemma and barriers in streaming and dynamic matching.
\newblock {\em CoRR}, abs/2207.09354, 2022.

\bibitem[ACC{\etalchar{+}}18]{ArarCCSW18}
Moab Arar, Shiri Chechik, Sarel Cohen, Cliff Stein, and David Wajc.
\newblock Dynamic matching: Reducing integral algorithms to
  approximately-maximal fractional algorithms.
\newblock In {\em {ICALP}}, volume 107 of {\em LIPIcs}, pages 7:1--7:16.
  Schloss Dagstuhl - Leibniz-Zentrum f{\"{u}}r Informatik, 2018.

\bibitem[AD16]{AbboudD16}
Amir Abboud and S{\o}ren Dahlgaard.
\newblock Popular conjectures as a barrier for dynamic planar graph algorithms.
\newblock In {\em {FOCS}}, pages 477--486. {IEEE} Computer Society, 2016.

\bibitem[AMV20]{AMV20}
Kyriakos Axiotis, Aleksander M{\k{a}}dry, and Adrian Vladu.
\newblock Circulation control for faster minimum cost flow in unit-capacity
  graphs.
\newblock In {\em 2020 IEEE 61st Annual Symposium on Foundations of Computer
  Science (FOCS)}, pages 93--104. IEEE, 2020.

\bibitem[AMV21]{AMV21}
Kyriakos Axiotis, Aleksander Madry, and Adrian Vladu.
\newblock Faster sparse minimum cost flow by electrical flow localization.
\newblock In {\em {FOCS}}, pages 528--539. {IEEE}, 2021.

\bibitem[AW14]{AbboudW14}
Amir Abboud and Virginia~Vassilevska Williams.
\newblock Popular conjectures imply strong lower bounds for dynamic problems.
\newblock In {\em {FOCS}}, pages 434--443. {IEEE} Computer Society, 2014.

\bibitem[BC18]{BC18}
Aaron Bernstein and Shiri Chechik.
\newblock Incremental topological sort and cycle detection in $o(m\sqrt{n})$
  expected total time.
\newblock In {\em {SODA}}, pages 21--34. {SIAM}, 2018.

\bibitem[BDH{\etalchar{+}}19]{BehnezhadDHSS19}
Soheil Behnezhad, Mahsa Derakhshan, MohammadTaghi Hajiaghayi, Cliff Stein, and
  Madhu Sudan.
\newblock Fully dynamic maximal independent set with polylogarithmic update
  time.
\newblock In {\em {FOCS}}, pages 382--405. {IEEE} Computer Society, 2019.

\bibitem[Beh22]{Behnezhad22}
Soheil Behnezhad.
\newblock Dynamic algorithms for maximum matching size.
\newblock {\em CoRR}, abs/2207.07607, 2022.

\bibitem[BFH21]{BernsteinFH21}
Aaron Bernstein, Sebastian Forster, and Monika Henzinger.
\newblock A deamortization approach for dynamic spanner and dynamic maximal
  matching.
\newblock {\em {ACM} Trans. Algorithms}, 17(4):29:1--29:51, 2021.

\bibitem[BGJ{\etalchar{+}}22]{BGJLLPS22}
Jan van~den Brand, Yu~Gao, Arun Jambulapati, Yin~Tat Lee, Yang~P. Liu, Richard
  Peng, and Aaron Sidford.
\newblock Faster maxflow via improved dynamic spectral vertex sparsifiers.
\newblock In {\em {STOC}}, pages 543--556. {ACM}, 2022.

\bibitem[BGS15]{BaswanaGS15}
Surender Baswana, Manoj Gupta, and Sandeep Sen.
\newblock Fully dynamic maximal matching in o(log n) update time.
\newblock {\em {SIAM} J. Comput.}, 44(1):88--113, 2015.

\bibitem[BGS21]{BGS21}
Aaron Bernstein, Maximilian~Probst Gutenberg, and Thatchaphol Saranurak.
\newblock Deterministic decremental {S}ssp and approximate min-cost flow in
  almost-linear time.
\newblock {\em arXiv preprint arXiv:2101.07149}, 2021.

\bibitem[BHN16]{BhattacharyaHN16}
Sayan Bhattacharya, Monika Henzinger, and Danupon Nanongkai.
\newblock New deterministic approximation algorithms for fully dynamic
  matching.
\newblock In {\em {STOC}}, pages 398--411. {ACM}, 2016.

\bibitem[BHR19]{BernsteinHR19}
Aaron Bernstein, Jacob Holm, and Eva Rotenberg.
\newblock Online bipartite matching with amortized \emph{O}(log
  \({}^{\mbox{2}}\) \emph{n}) replacements.
\newblock {\em J. {ACM}}, 66(5):37:1--37:23, 2019.

\bibitem[BK19]{BhattacharyaK19}
Sayan Bhattacharya and Janardhan Kulkarni.
\newblock Deterministically maintaining a $(2+\epsilon)$-approximate minimum
  vertex cover in $o(1/\epsilon^2)$ amortized update time.
\newblock In {\em {SODA}}, pages 1872--1885. {SIAM}, 2019.

\bibitem[BK20]{BK20}
Sayan Bhattacharya and Janardhan Kulkarni.
\newblock An improved algorithm for incremental cycle detection and topological
  ordering in sparse graphs.
\newblock In {\em {SODA}}, pages 2509--2521. {SIAM}, 2020.

\bibitem[BK21]{BhattacharyaK21}
Sayan Bhattacharya and Peter Kiss.
\newblock Deterministic rounding of dynamic fractional matchings.
\newblock In {\em {ICALP}}, volume 198 of {\em LIPIcs}, pages 27:1--27:14.
  Schloss Dagstuhl - Leibniz-Zentrum f{\"{u}}r Informatik, 2021.

\bibitem[BK22]{BehnezhadK22}
Soheil Behnezhad and Sanjeev Khanna.
\newblock New trade-offs for fully dynamic matching via hierarchical {EDCS}.
\newblock In {\em {SODA}}, pages 3529--3566. {SIAM}, 2022.

\bibitem[BKS22]{BKS22}
Sayan Bhattacharya, Peter Kiss, and Thatchaphol Saranurak.
\newblock Dynamic algorithms for packing-covering lps via multiplicative weight
  updates.
\newblock {\em arXiv preprint arXiv:2207.07519}, 2022.

\bibitem[BKSW22]{BhattacharyaKSW22}
Sayan Bhattacharya, Peter Kiss, Thatchaphol Saranurak, and David Wajc.
\newblock Dynamic matching with better-than-2 approximation in polylogarithmic
  update time.
\newblock {\em CoRR}, abs/2207.07438, 2022.

\bibitem[BLL{\etalchar{+}}21]{BLLSSSW21}
Jan van~den Brand, Yin~Tat Lee, Yang~P. Liu, Thatchaphol Saranurak, Aaron
  Sidford, Zhao Song, and Di~Wang.
\newblock Minimum cost flows, mdps, and $\ell_1$-regression in nearly linear
  time for dense instances.
\newblock In {\em {STOC}}, pages 859--869. {ACM}, 2021.

\bibitem[BLM20]{BehnezhadLM20}
Soheil Behnezhad, Jakub Lacki, and Vahab~S. Mirrokni.
\newblock Fully dynamic matching: Beating 2-approximation in $\delta^\epsilon$
  update time.
\newblock In {\em {SODA}}, pages 2492--2508. {SIAM}, 2020.

\bibitem[BLN{\etalchar{+}}20]{BLNPSSSW20}
Jan van~den Brand, Yin-Tat Lee, Danupon Nanongkai, Richard Peng, Thatchaphol
  Saranurak, Aaron Sidford, Zhao Song, and Di~Wang.
\newblock Bipartite matching in nearly-linear time on moderately dense graphs.
\newblock In {\em 2020 IEEE 61st Annual Symposium on Foundations of Computer
  Science (FOCS)}, pages 919--930. IEEE, 2020.

\bibitem[BLSZ14]{BosekLSZ14}
Bartlomiej Bosek, Dariusz Leniowski, Piotr Sankowski, and Anna Zych.
\newblock Online bipartite matching in offline time.
\newblock In {\em {FOCS}}, pages 384--393. {IEEE} Computer Society, 2014.

\bibitem[BNS19]{BrandNS19}
Jan van~den Brand, Danupon Nanongkai, and Thatchaphol Saranurak.
\newblock Dynamic matrix inverse: Improved algorithms and matching conditional
  lower bounds.
\newblock In {\em {FOCS}}, pages 456--480. {IEEE} Computer Society, 2019.

\bibitem[BS15]{BernsteinS15}
Aaron Bernstein and Cliff Stein.
\newblock Fully dynamic matching in bipartite graphs.
\newblock In {\em {ICALP} {(1)}}, volume 9134 of {\em Lecture Notes in Computer
  Science}, pages 167--179. Springer, 2015.

\bibitem[BS16]{BernsteinS16}
Aaron Bernstein and Cliff Stein.
\newblock Faster fully dynamic matchings with small approximation ratios.
\newblock In {\em {SODA}}, pages 692--711. {SIAM}, 2016.

\bibitem[CGH{\etalchar{+}}20]{CGHPS20}
Li~Chen, Gramoz Goranci, Monika Henzinger, Richard Peng, and Thatchaphol
  Saranurak.
\newblock Fast dynamic cuts, distances and effective resistances via vertex
  sparsifiers.
\newblock In {\em 2020 IEEE 61st Annual Symposium on Foundations of Computer
  Science (FOCS)}, pages 1135--1146. IEEE, 2020.

\bibitem[CGL{\etalchar{+}}20]{CGLNPS20}
Julia Chuzhoy, Yu~Gao, Jason Li, Danupon Nanongkai, Richard Peng, and
  Thatchaphol Saranurak.
\newblock A deterministic algorithm for balanced cut with applications to
  dynamic connectivity, flows, and beyond.
\newblock In {\em {FOCS}}, pages 1158--1167. {IEEE}, 2020.

\bibitem[CKL{\etalchar{+}}22]{CKLPPS22}
Li~Chen, Rasmus Kyng, Yang~P Liu, Richard Peng, Maximilian~Probst Gutenberg,
  and Sushant Sachdeva.
\newblock Maximum flow and minimum-cost flow in almost-linear time.
\newblock {\em arXiv preprint arXiv:2203.00671}, 2022.

\bibitem[CMSV17]{CMSV17}
Michael~B. Cohen, Aleksander M\k{a}dry, Piotr Sankowski, and Adrian Vladu.
\newblock Negative-weight shortest paths and unit capacity minimum cost flow in
  $\widetilde{O}(m^{10/7})$ time (extended abstract).
\newblock In {\em {SODA}}, pages 752--771. {SIAM}, 2017.

\bibitem[CS18]{CharikarS18}
Moses Charikar and Shay Solomon.
\newblock Fully dynamic almost-maximal matching: Breaking the polynomial
  worst-case time barrier.
\newblock In {\em {ICALP}}, volume 107 of {\em LIPIcs}, pages 33:1--33:14.
  Schloss Dagstuhl - Leibniz-Zentrum f{\"{u}}r Informatik, 2018.

\bibitem[CZ19]{ChechikZ19}
Shiri Chechik and Tianyi Zhang.
\newblock Fully dynamic maximal independent set in expected poly-log update
  time.
\newblock In {\em {FOCS}}, pages 370--381. {IEEE} Computer Society, 2019.

\bibitem[Dah16]{Dahlgaard16}
S{\o}ren Dahlgaard.
\newblock On the hardness of partially dynamic graph problems and connections
  to diameter.
\newblock In {\em {ICALP}}, volume~55 of {\em LIPIcs}, pages 48:1--48:14.
  Schloss Dagstuhl - Leibniz-Zentrum f{\"{u}}r Informatik, 2016.

\bibitem[DGG{\etalchar{+}}22]{DGGLPSY22}
Sally Dong, Yu~Gao, Gramoz Goranci, Yin~Tat Lee, Richard Peng, Sushant
  Sachdeva, and Guanghao Ye.
\newblock Nested dissection meets ipms: Planar min-cost flow in nearly-linear
  time.
\newblock In {\em {SODA}}, pages 124--153. {SIAM}, 2022.

\bibitem[EGIN97]{EGIN97}
David Eppstein, Zvi Galil, Giuseppe~F. Italiano, and Amnon Nissenzweig.
\newblock Sparsification - a technique for speeding up dynamic graph
  algorithms.
\newblock {\em J. {ACM}}, 44(5):669--696, 1997.

\bibitem[ET75]{ET75}
Shimon Even and R.~Endre Tarjan.
\newblock Network flow and testing graph connectivity.
\newblock {\em SIAM journal on computing}, 4(4):507--518, 1975.

\bibitem[Fre85]{Fred85}
Greg~N. Frederickson.
\newblock Data structures for on-line updating of minimum spanning trees, with
  applications.
\newblock {\em {SIAM} J. Comput.}, 14(4):781--798, 1985.

\bibitem[GH22]{Goranci}
Gramoz Goranci and Monika Henzinger.
\newblock Incremental approximate maximum flow in $m^{1/2+o(1)}$ update time.
\newblock {\em arXiv preprint arXiv:2211.09606}, 2022.
\newblock Available at \url{https://arxiv.org/pdf/2211.09606.pdf}.

\bibitem[GK07]{GK07}
Naveen Garg and Jochen K{\"{o}}nemann.
\newblock Faster and simpler algorithms for multicommodity flow and other
  fractional packing problems.
\newblock {\em {SIAM} J. Comput.}, 37(2):630--652, 2007.

\bibitem[GK21]{GuptaK21}
Manoj Gupta and Shahbaz Khan.
\newblock Simple dynamic algorithms for maximal independent set, maximum flow
  and maximum matching.
\newblock In {\em {SOSA}}, pages 86--91. {SIAM}, 2021.

\bibitem[GLP21]{GLP21}
Yu~Gao, Yang~P. Liu, and Richard Peng.
\newblock Fully dynamic electrical flows: Sparse maxflow faster than
  goldberg-rao.
\newblock In {\em {FOCS}}, pages 516--527. {IEEE}, 2021.

\bibitem[GP13]{GuptaP13}
Manoj Gupta and Richard Peng.
\newblock Fully dynamic {(1+} e)-approximate matchings.
\newblock In {\em {FOCS}}, pages 548--557. {IEEE} Computer Society, 2013.

\bibitem[GR98]{GR98}
Andrew~V. Goldberg and Satish Rao.
\newblock Beyond the flow decomposition barrier.
\newblock {\em Journal of the ACM}, 45(5):783--797, 1998.
\newblock Announced at FOCS'97.

\bibitem[GRST21]{GRST21}
Gramoz Goranci, Harald R{\"{a}}cke, Thatchaphol Saranurak, and Zihan Tan.
\newblock The expander hierarchy and its applications to dynamic graph
  algorithms.
\newblock In {\em {SODA}}, pages 2212--2228. {SIAM}, 2021.

\bibitem[GSSU22]{GrandoniSSU22}
Fabrizio Grandoni, Chris Schwiegelshohn, Shay Solomon, and Amitai Uzrad.
\newblock Maintaining an {EDCS} in general graphs: Simpler, density-sensitive
  and with worst-case time bounds.
\newblock In {\em {SOSA}}, pages 12--23. {SIAM}, 2022.

\bibitem[Gup14]{Gupta14}
Manoj Gupta.
\newblock Maintaining approximate maximum matching in an incremental bipartite
  graph in polylogarithmic update time.
\newblock In {\em {FSTTCS}}, volume~29 of {\em LIPIcs}, pages 227--239. Schloss
  Dagstuhl - Leibniz-Zentrum f{\"{u}}r Informatik, 2014.

\bibitem[HdLT01]{HLT01}
Jacob Holm, Kristian de~Lichtenberg, and Mikkel Thorup.
\newblock Poly-logarithmic deterministic fully-dynamic algorithms for
  connectivity, minimum spanning tree, 2-edge, and biconnectivity.
\newblock {\em J. {ACM}}, 48(4):723--760, 2001.

\bibitem[HHKP17]{HHKP17}
Shang{-}En Huang, Dawei Huang, Tsvi Kopelowitz, and Seth Pettie.
\newblock Fully dynamic connectivity in $o(\log n (\log \log n)^2)$ amortized
  expected time.
\newblock In {\em {SODA}}, pages 510--520. {SIAM}, 2017.

\bibitem[HK97]{HK97}
Monika~Rauch Henzinger and Valerie King.
\newblock Maintaining minimum spanning trees in dynamic graphs.
\newblock In {\em {ICALP}}, volume 1256 of {\em Lecture Notes in Computer
  Science}, pages 594--604. Springer, 1997.

\bibitem[HK99]{HK99}
Monika~Rauch Henzinger and Valerie King.
\newblock Randomized fully dynamic graph algorithms with polylogarithmic time
  per operation.
\newblock {\em J. {ACM}}, 46(4):502--516, 1999.

\bibitem[HKNS15]{HenzingerKNS15}
Monika Henzinger, Sebastian Krinninger, Danupon Nanongkai, and Thatchaphol
  Saranurak.
\newblock Unifying and strengthening hardness for dynamic problems via the
  online matrix-vector multiplication conjecture.
\newblock In {\em {STOC}}, pages 21--30. {ACM}, 2015.

\bibitem[HRW15]{HRW15}
Jacob Holm, Eva Rotenberg, and Christian Wulff{-}Nilsen.
\newblock Faster fully-dynamic minimum spanning forest.
\newblock In {\em {ESA}}, volume 9294 of {\em Lecture Notes in Computer
  Science}, pages 742--753. Springer, 2015.

\bibitem[HT97]{HT97}
Monika~Rauch Henzinger and Mikkel Thorup.
\newblock Sampling to provide or to bound: With applications to fully dynamic
  graph algorithms.
\newblock {\em Random Struct. Algorithms}, 11(4):369--379, 1997.

\bibitem[Ita86]{Italiano86}
Giuseppe~F. Italiano.
\newblock Amortized efficiency of a path retrieval data structure.
\newblock {\em Theor. Comput. Sci.}, 48(3):273--281, 1986.

\bibitem[JJST22]{JJST22}
Arun Jambulapati, Yujia Jin, Aaron Sidford, and Kevin Tian.
\newblock Regularized box-simplex games and dynamic decremental bipartite
  matching.
\newblock In {\em {ICALP}}, volume 229 of {\em LIPIcs}, pages 77:1--77:20.
  Schloss Dagstuhl - Leibniz-Zentrum f{\"{u}}r Informatik, 2022.

\bibitem[Kar73]{K73}
Alexander~V Karzanov.
\newblock On finding maximum flows in networks with special structure and some
  applications.
\newblock {\em Matematicheskie Voprosy Upravleniya Proizvodstvom}, 5:81--94,
  1973.

\bibitem[KG03]{KumarG03}
S.~Kumar and P.~Gupta.
\newblock An incremental algorithm for the maximum flow problem.
\newblock {\em J. Math. Model. Algorithms}, 2(1):1--16, 2003.

\bibitem[Kis21]{Kiss21}
Peter Kiss.
\newblock Improving update times of dynamic matching algorithms from amortized
  to worst case.
\newblock {\em CoRR}, abs/2108.10461, 2021.

\bibitem[KLOS14]{KLOS14}
Jonathan~A. Kelner, Yin~Tat Lee, Lorenzo Orecchia, and Aaron Sidford.
\newblock An almost-linear-time algorithm for approximate max flow in
  undirected graphs, and its multicommodity generalizations.
\newblock In Chandra Chekuri, editor, {\em Proceedings of the Twenty-Fifth
  Annual {ACM-SIAM} Symposium on Discrete Algorithms, {SODA} 2014, Portland,
  Oregon, USA, January 5-7, 2014}, pages 217--226. {SIAM}, 2014.

\bibitem[KLS20]{KLS20}
Tarun Kathuria, Yang~P. Liu, and Aaron Sidford.
\newblock Unit capacity maxflow in almost {$O(m^{4/3})$} time.
\newblock In {\em 61st {IEEE} Annual Symposium on Foundations of Computer
  Science, {FOCS} 2020, Durham, NC, USA, November 16-19, 2020}, pages 119--130.
  {IEEE}, 2020.

\bibitem[KPP16]{KopelowitzPP16}
Tsvi Kopelowitz, Seth Pettie, and Ely Porat.
\newblock Higher lower bounds from the 3sum conjecture.
\newblock In {\em {SODA}}, pages 1272--1287. {SIAM}, 2016.

\bibitem[LMSW22]{LeMSW22}
Hung Le, Lazar Milenkovic, Shay Solomon, and Virginia~Vassilevska Williams.
\newblock Dynamic matching algorithms under vertex updates.
\newblock In {\em {ITCS}}, volume 215 of {\em LIPIcs}, pages 96:1--96:24.
  Schloss Dagstuhl - Leibniz-Zentrum f{\"{u}}r Informatik, 2022.

\bibitem[LS14]{LS14}
Yin~Tat Lee and Aaron Sidford.
\newblock Path finding methods for linear programming: Solving linear programs
  in $\o(\sqrt{\mathrm{rank}})$ iterations and faster algorithms for maximum
  flow.
\newblock In {\em {FOCS}}, pages 424--433. {IEEE} Computer Society, 2014.

\bibitem[LS20]{LS20}
Yang~P Liu and Aaron Sidford.
\newblock Faster energy maximization for faster maximum flow.
\newblock In {\em Proceedings of the 52nd Annual ACM SIGACT Symposium on Theory
  of Computing}, pages 803--814, 2020.

\bibitem[M{\k{a}}d13]{M13}
Aleksander M{\k{a}}dry.
\newblock Navigating central path with electrical flows: From flows to
  matchings, and back.
\newblock In {\em 2013 IEEE 54th Annual Symposium on Foundations of Computer
  Science}, pages 253--262. IEEE, 2013.

\bibitem[M{\k{a}}d16]{M16}
Aleksander M{\k{a}}dry.
\newblock Computing maximum flow with augmenting electrical flows.
\newblock In {\em 57th {IEEE} Annual Symposium on Foundations of Computer
  Science, {FOCS} 2016, 9-11 October 2016, Hyatt Regency, New Brunswick, New
  Jersey, {USA}}, pages 593--602. {IEEE} Computer Society, 2016.
\newblock Available at~\url{https://arxiv.org/abs/1608.06016}.

\bibitem[NS17]{NS17}
Danupon Nanongkai and Thatchaphol Saranurak.
\newblock Dynamic spanning forest with worst-case update time: adaptive, las
  vegas, and $o(n^{1/2-\epsilon})$-time.
\newblock In {\em {STOC}}, pages 1122--1129. {ACM}, 2017.

\bibitem[NSW17]{NSW17}
Danupon Nanongkai, Thatchaphol Saranurak, and Christian Wulff{-}Nilsen.
\newblock Dynamic minimum spanning forest with subpolynomial worst-case update
  time.
\newblock In Chris Umans, editor, {\em 58th {IEEE} Annual Symposium on
  Foundations of Computer Science, {FOCS} 2017, Berkeley, CA, USA, October
  15-17, 2017}, pages 950--961. {IEEE} Computer Society, 2017.
\newblock Available at:~\url{https://arxiv.org/abs/1708.03962}.

\bibitem[PD06]{PD06}
Mihai Patrascu and Erik~D. Demaine.
\newblock Logarithmic lower bounds in the cell-probe model.
\newblock {\em {SIAM} J. Comput.}, 35(4):932--963, 2006.

\bibitem[Pen16]{P16}
Richard Peng.
\newblock Approximate undirected maximum flows in $o(m \mathrm{polylog}(n))$
  time.
\newblock In {\em Proceedings of the twenty-seventh annual ACM-SIAM symposium
  on Discrete algorithms}, pages 1862--1867. SIAM, 2016.

\bibitem[PS16]{PelegS16}
David Peleg and Shay Solomon.
\newblock Dynamic $(1 + \epsilon)$-approximate matchings: {A} density-sensitive
  approach.
\newblock In {\em {SODA}}, pages 712--729. {SIAM}, 2016.

\bibitem[R{\"{a}}c08]{Racke08}
Harald R{\"{a}}cke.
\newblock Optimal hierarchical decompositions for congestion minimization in
  networks.
\newblock In {\em {STOC}}, pages 255--264. {ACM}, 2008.

\bibitem[RSW22]{RoghaniSW22}
Mohammad Roghani, Amin Saberi, and David Wajc.
\newblock Beating the folklore algorithm for dynamic matching.
\newblock In {\em {ITCS}}, volume 215 of {\em LIPIcs}, pages 111:1--111:23.
  Schloss Dagstuhl - Leibniz-Zentrum f{\"{u}}r Informatik, 2022.

\bibitem[San07]{Sankowski07}
Piotr Sankowski.
\newblock Faster dynamic matchings and vertex connectivity.
\newblock In {\em {SODA}}, pages 118--126. {SIAM}, 2007.

\bibitem[She13]{S13}
Jonah Sherman.
\newblock Nearly maximum flows in nearly linear time.
\newblock In {\em 54th Annual {IEEE} Symposium on Foundations of Computer
  Science, {FOCS} 2013, 26-29 October, 2013, Berkeley, CA, {USA}}, pages
  263--269. {IEEE} Computer Society, 2013.

\bibitem[Sol16]{Solomon16}
Shay Solomon.
\newblock Fully dynamic maximal matching in constant update time.
\newblock In {\em {FOCS}}, pages 325--334. {IEEE} Computer Society, 2016.

\bibitem[Tho00]{Thorup00}
Mikkel Thorup.
\newblock Near-optimal fully-dynamic graph connectivity.
\newblock In {\em {STOC}}, pages 343--350. {ACM}, 2000.

\bibitem[Waj20]{Wajc20}
David Wajc.
\newblock Rounding dynamic matchings against an adaptive adversary.
\newblock In {\em {STOC}}, pages 194--207. {ACM}, 2020.

\bibitem[Wul13]{WN13a}
Christian Wulff{-}Nilsen.
\newblock Faster deterministic fully-dynamic graph connectivity.
\newblock In {\em {SODA}}, pages 1757--1769. {SIAM}, 2013.

\bibitem[Wul17]{WN17}
Christian Wulff{-}Nilsen.
\newblock Fully-dynamic minimum spanning forest with improved worst-case update
  time.
\newblock In {\em {STOC}}, pages 1130--1143. {ACM}, 2017.

\end{thebibliography}

%\appendix

%\input{proofs}

\end{document}